\newcount\who 
\documentstyle[12pt]{article}
\ifx\undefined\psfig\else\endinput\fi

\edef\psfigRestoreAt{\catcode`@=\number\catcode`@\relax}
\catcode`\@=11\relax
\newwrite\@unused
\def\ps@typeout#1{{\let\protect\string\immediate\write\@unused{#1}}}
\ps@typeout{psfig/tex 1.8}

\def\figurepath{./}

\def\@nnil{\@nil}
\def\@empty{}
\def\@psdonoop#1\@@#2#3{}
\def\@psdo#1:=#2\do#3{\edef\@psdotmp{#2}\ifx\@psdotmp\@empty \else
    \expandafter\@psdoloop#2,\@nil,\@nil\@@#1{#3}\fi}
\def\@psdoloop#1,#2,#3\@@#4#5{\def#4{#1}\ifx #4\@nnil \else
       #5\def#4{#2}\ifx #4\@nnil \else#5\@ipsdoloop #3\@@#4{#5}\fi\fi}
\def\@ipsdoloop#1,#2\@@#3#4{\def#3{#1}\ifx #3\@nnil 
       \let\@nextwhile=\@psdonoop \else
      #4\relax\let\@nextwhile=\@ipsdoloop\fi\@nextwhile#2\@@#3{#4}}
\def\@tpsdo#1:=#2\do#3{\xdef\@psdotmp{#2}\ifx\@psdotmp\@empty \else
    \@tpsdoloop#2\@nil\@nil\@@#1{#3}\fi}
\def\@tpsdoloop#1#2\@@#3#4{\def#3{#1}\ifx #3\@nnil 
       \let\@nextwhile=\@psdonoop \else
      #4\relax\let\@nextwhile=\@tpsdoloop\fi\@nextwhile#2\@@#3{#4}}
\ifx\undefined\fbox
\newdimen\fboxrule
\newdimen\fboxsep
\newdimen\ps@tempdima
\newbox\ps@tempboxa
\long\def\fbox#1{\leavevmode\setbox\ps@tempboxa\hbox{#1}\ps@tempdima\fboxrule
    \advance\ps@tempdima \fboxsep \advance\ps@tempdima \dp\ps@tempboxa
   \hbox{\lower \ps@tempdima\hbox
  {\vbox{\hrule height \fboxrule
          \hbox{\vrule width \fboxrule \hskip\fboxsep
          \vbox{\vskip\fboxsep \box\ps@tempboxa\vskip\fboxsep}\hskip 
                 \fboxsep\vrule width \fboxrule}
                 \hrule height \fboxrule}}}}
\fi
\newread\ps@stream
\newif\ifnot@eof       
\newif\if@noisy        
\newif\if@atend        
\newif\if@psfile       
{\catcode`\%=12\global\gdef\epsf@start{
\def\epsf@PS{PS}
\def\epsf@getbb#1{%
\openin\ps@stream=#1
\ifeof\ps@stream\ps@typeout{Error, File #1 not found}\else
   {\not@eoftrue \chardef\other=12
    \def\do##1{\catcode`##1=\other}\dospecials \catcode`\ =10
    \loop
       \if@psfile
	  \read\ps@stream to \epsf@fileline
       \else{
	  \obeyspaces
          \read\ps@stream to \epsf@tmp\global\let\epsf@fileline\epsf@tmp}
       \fi
       \ifeof\ps@stream\not@eoffalse\else
       \if@psfile\else
       \expandafter\epsf@test\epsf@fileline:. \\%
       \fi
          \expandafter\epsf@aux\epsf@fileline:. \\%
       \fi
   \ifnot@eof\repeat
   }\closein\ps@stream\fi}%
\long\def\epsf@test#1#2#3:#4\\{\def\epsf@testit{#1#2}
			\ifx\epsf@testit\epsf@start\else
\ps@typeout{Warning! File does not start with `\epsf@start'.  It may not be a PostScript file.}
			\fi
			\@psfiletrue} 
{\catcode`\%=12\global\let\epsf@percent=
\long\def\epsf@aux#1#2:#3\\{\ifx#1\epsf@percent
   \def\epsf@testit{#2}\ifx\epsf@testit\epsf@bblit
	\@atendfalse
        \epsf@atend #3 . \\%
	\if@atend	
	   \if@verbose{
		\ps@typeout{psfig: found `(atend)'; continuing search}
	   }\fi
        \else
        \epsf@grab #3 . . . \\%
        \not@eoffalse
        \global\no@bbfalse
        \fi
   \fi\fi}%
\def\epsf@grab #1 #2 #3 #4 #5\\{%
   \global\def\epsf@llx{#1}\ifx\epsf@llx\empty
      \epsf@grab #2 #3 #4 #5 .\\\else
   \global\def\epsf@lly{#2}%
   \global\def\epsf@urx{#3}\global\def\epsf@ury{#4}\fi}%
\def\epsf@atendlit{(atend)} 
\def\epsf@atend #1 #2 #3\\{%
   \def\epsf@tmp{#1}\ifx\epsf@tmp\empty
      \epsf@atend #2 #3 .\\\else
   \ifx\epsf@tmp\epsf@atendlit\@atendtrue\fi\fi}

\chardef\letter = 11
\chardef\other = 12

\newif \ifdebug 
\newif\ifc@mpute 
\c@mputetrue 

\let\then = \relax
\def\r@dian{pt }
\let\r@dians = \r@dian
\let\dimensionless@nit = \r@dian
\let\dimensionless@nits = \dimensionless@nit
\def\internal@nit{sp }
\let\internal@nits = \internal@nit
\newif\ifstillc@nverging
\def \Mess@ge #1{\ifdebug \then \message {#1} \fi}

{ 
	\catcode `\@ = \letter
	\gdef \nodimen {\expandafter \n@dimen \the \dimen}
	\gdef \term #1 #2 #3%
	       {\edef \t@ {\the #1}
		\edef \t@@ {\expandafter \n@dimen \the #2\r@dian}%
		\t@rm {\t@} {\t@@} {#3}%
	       }
	\gdef \t@rm #1 #2 #3%
	       {{%
		\count 0 = 0
		\dimen 0 = 1 \dimensionless@nit
		\dimen 2 = #2\relax
		\Mess@ge {Calculating term #1 of \nodimen 2}%
		\loop
		\ifnum	\count 0 < #1
		\then	\advance \count 0 by 1
			\Mess@ge {Iteration \the \count 0 \space}%
			\Multiply \dimen 0 by {\dimen 2}%
			\Mess@ge {After multiplication, term = \nodimen 0}%
			\Divide \dimen 0 by {\count 0}%
			\Mess@ge {After division, term = \nodimen 0}%
		\repeat
		\Mess@ge {Final value for term #1 of 
				\nodimen 2 \space is \nodimen 0}%
		\xdef \Term {#3 = \nodimen 0 \r@dians}%
		\aftergroup \Term
	       }}
	\catcode `\p = \other
	\catcode `\t = \other
	\gdef \n@dimen #1pt{#1} 
}

\def \Divide #1by #2{\divide #1 by #2} 

\def \Multiply #1by #2
       {{
	\count 0 = #1\relax
	\count 2 = #2\relax
	\count 4 = 65536
	\Mess@ge {Before scaling, count 0 = \the \count 0 \space and
			count 2 = \the \count 2}%
	\ifnum	\count 0 > 32767 
	\then	\divide \count 0 by 4
		\divide \count 4 by 4
	\else	\ifnum	\count 0 < -32767
		\then	\divide \count 0 by 4
			\divide \count 4 by 4
		\else
		\fi
	\fi
	\ifnum	\count 2 > 32767 
	\then	\divide \count 2 by 4
		\divide \count 4 by 4
	\else	\ifnum	\count 2 < -32767
		\then	\divide \count 2 by 4
			\divide \count 4 by 4
		\else
		\fi
	\fi
	\multiply \count 0 by \count 2
	\divide \count 0 by \count 4
	\xdef \product {#1 = \the \count 0 \internal@nits}%
	\aftergroup \product
       }}

\def\r@duce{\ifdim\dimen0 > 90\r@dian \then   
		\multiply\dimen0 by -1
		\advance\dimen0 by 180\r@dian
		\r@duce
	    \else \ifdim\dimen0 < -90\r@dian \then  
		\advance\dimen0 by 360\r@dian
		\r@duce
		\fi
	    \fi}

\def\Sine#1%
       {{%
	\dimen 0 = #1 \r@dian
	\r@duce
	\ifdim\dimen0 = -90\r@dian \then
	   \dimen4 = -1\r@dian
	   \c@mputefalse
	\fi
	\ifdim\dimen0 = 90\r@dian \then
	   \dimen4 = 1\r@dian
	   \c@mputefalse
	\fi
	\ifdim\dimen0 = 0\r@dian \then
	   \dimen4 = 0\r@dian
	   \c@mputefalse
	\fi
	\ifc@mpute \then
		\divide\dimen0 by 180
		\dimen0=3.141592654\dimen0
		\dimen 2 = 3.1415926535897963\r@dian 
		\divide\dimen 2 by 2 
		\Mess@ge {Sin: calculating Sin of \nodimen 0}%
		\count 0 = 1 
		\dimen 2 = 1 \r@dian 
		\dimen 4 = 0 \r@dian 
		\loop
			\ifnum	\dimen 2 = 0 
			\then	\stillc@nvergingfalse 
			\else	\stillc@nvergingtrue
			\fi
			\ifstillc@nverging 
			\then	\term {\count 0} {\dimen 0} {\dimen 2}%
				\advance \count 0 by 2
				\count 2 = \count 0
				\divide \count 2 by 2
				\ifodd	\count 2 
				\then	\advance \dimen 4 by \dimen 2
				\else	\advance \dimen 4 by -\dimen 2
				\fi
		\repeat
	\fi		
			\xdef \sine {\nodimen 4}%
       }}

\def\Cosine#1{\ifx\sine\UnDefined\edef\Savesine{\relax}\else
		             \edef\Savesine{\sine}\fi
	{\dimen0=#1\r@dian\advance\dimen0 by 90\r@dian
	 \Sine{\nodimen 0}
	 \xdef\cosine{\sine}
	 \xdef\sine{\Savesine}}}	      

\def\psdraft{
	\def\@psdraft{0}
}
\def\psfull{
	\def\@psdraft{100}
}

\psfull

\newif\if@scalefirst
\def\psscalefirst{\@scalefirsttrue}
\def\psrotatefirst{\@scalefirstfalse}
\psrotatefirst

\newif\if@draftbox
\def\psnodraftbox{
	\@draftboxfalse
}
\def\psdraftbox{
	\@draftboxtrue
}
\@draftboxtrue

\newif\if@prologfile
\newif\if@postlogfile
\def\pssilent{
	\@noisyfalse
}
\def\psnoisy{
	\@noisytrue
}
\psnoisy
\newif\if@bbllx
\newif\if@bblly
\newif\if@bburx
\newif\if@bbury
\newif\if@height
\newif\if@width
\newif\if@rheight
\newif\if@rwidth
\newif\if@angle
\newif\if@clip
\newif\if@verbose
\def\@p@@sclip#1{\@cliptrue}

\newif\if@decmpr

\def\@p@@sfigure#1{\def\@p@sfile{null}\def\@p@sbbfile{null}
	        \openin1=#1.bb
		\ifeof1\closein1
	        	\openin1=\figurepath#1.bb
			\ifeof1\closein1
			        \openin1=#1
				\ifeof1\closein1%
				       \openin1=\figurepath#1
					\ifeof1
					   \ps@typeout{Error, File #1 not found}
						\if@bbllx\if@bblly
				   		\if@bburx\if@bbury
			      				\def\@p@sfile{#1}%
			      				\def\@p@sbbfile{#1}%
							\@decmprfalse
				  	   	\fi\fi\fi\fi
					\else\closein1
				    		\def\@p@sfile{\figurepath#1}%
				    		\def\@p@sbbfile{\figurepath#1}%
						\@decmprfalse
	                       		\fi%
			 	\else\closein1%
					\def\@p@sfile{#1}
					\def\@p@sbbfile{#1}
					\@decmprfalse
			 	\fi
			\else
				\def\@p@sfile{\figurepath#1}
				\def\@p@sbbfile{\figurepath#1.bb}
				\@decmprtrue
			\fi
		\else
			\def\@p@sfile{#1}
			\def\@p@sbbfile{#1.bb}
			\@decmprtrue
		\fi}

\def\@p@@sfile#1{\@p@@sfigure{#1}}

\def\@p@@sbbllx#1{
		\@bbllxtrue
		\dimen100=#1
		\edef\@p@sbbllx{\number\dimen100}
}
\def\@p@@sbblly#1{
		\@bbllytrue
		\dimen100=#1
		\edef\@p@sbblly{\number\dimen100}
}
\def\@p@@sbburx#1{
		\@bburxtrue
		\dimen100=#1
		\edef\@p@sbburx{\number\dimen100}
}
\def\@p@@sbbury#1{
		\@bburytrue
		\dimen100=#1
		\edef\@p@sbbury{\number\dimen100}
}
\def\@p@@sheight#1{
		\@heighttrue
		\dimen100=#1
   		\edef\@p@sheight{\number\dimen100}
}
\def\@p@@swidth#1{
		\@widthtrue
		\dimen100=#1
		\edef\@p@swidth{\number\dimen100}
}
\def\@p@@srheight#1{
		\@rheighttrue
		\dimen100=#1
		\edef\@p@srheight{\number\dimen100}
}
\def\@p@@srwidth#1{
		\@rwidthtrue
		\dimen100=#1
		\edef\@p@srwidth{\number\dimen100}
}
\def\@p@@sangle#1{
		\@angletrue
		\edef\@p@sangle{#1} 
}
\def\@p@@ssilent#1{ 
		\@verbosefalse
}
\def\@p@@sprolog#1{\@prologfiletrue\def\@prologfileval{#1}}
\def\@p@@spostlog#1{\@postlogfiletrue\def\@postlogfileval{#1}}
\def\@cs@name#1{\csname #1\endcsname}
\def\@setparms#1=#2,{\@cs@name{@p@@s#1}{#2}}
\def\ps@init@parms{
		\@bbllxfalse \@bbllyfalse
		\@bburxfalse \@bburyfalse
		\@heightfalse \@widthfalse
		\@rheightfalse \@rwidthfalse
		\def\@p@sbbllx{}\def\@p@sbblly{}
		\def\@p@sbburx{}\def\@p@sbbury{}
		\def\@p@sheight{}\def\@p@swidth{}
		\def\@p@srheight{}\def\@p@srwidth{}
		\def\@p@sangle{0}
		\def\@p@sfile{} \def\@p@sbbfile{}
		\def\@p@scost{10}
		\def\@sc{}
		\@prologfilefalse
		\@postlogfilefalse
		\@clipfalse
		\if@noisy
			\@verbosetrue
		\else
			\@verbosefalse
		\fi
}
\def\parse@ps@parms#1{
	 	\@psdo\@psfiga:=#1\do
		   {\expandafter\@setparms\@psfiga,}}
\newif\ifno@bb
\def\bb@missing{
	\if@verbose{
		\ps@typeout{psfig: searching \@p@sbbfile \space  for bounding box}
	}\fi
	\no@bbtrue
	\epsf@getbb{\@p@sbbfile}
        \ifno@bb \else \bb@cull\epsf@llx\epsf@lly\epsf@urx\epsf@ury\fi
}	
\def\bb@cull#1#2#3#4{
	\dimen100=#1 bp\edef\@p@sbbllx{\number\dimen100}
	\dimen100=#2 bp\edef\@p@sbblly{\number\dimen100}
	\dimen100=#3 bp\edef\@p@sbburx{\number\dimen100}
	\dimen100=#4 bp\edef\@p@sbbury{\number\dimen100}
	\no@bbfalse
}
\newdimen\p@intvaluex
\newdimen\p@intvaluey
\def\rotate@#1#2{{\dimen0=#1 sp\dimen1=#2 sp
		  \global\p@intvaluex=\cosine\dimen0
		  \dimen3=\sine\dimen1
		  \global\advance\p@intvaluex by -\dimen3
		  \global\p@intvaluey=\sine\dimen0
		  \dimen3=\cosine\dimen1
		  \global\advance\p@intvaluey by \dimen3
		  }}
\def\compute@bb{
		\no@bbfalse
		\if@bbllx \else \no@bbtrue \fi
		\if@bblly \else \no@bbtrue \fi
		\if@bburx \else \no@bbtrue \fi
		\if@bbury \else \no@bbtrue \fi
		\ifno@bb \bb@missing \fi
		\ifno@bb \ps@typeout{FATAL ERROR: no bb supplied or found}
			\no-bb-error
		\fi
		\count203=\@p@sbburx
		\count204=\@p@sbbury
		\advance\count203 by -\@p@sbbllx
		\advance\count204 by -\@p@sbblly
		\edef\ps@bbw{\number\count203}
		\edef\ps@bbh{\number\count204}
		\if@angle 
			\Sine{\@p@sangle}\Cosine{\@p@sangle}
	        	{\dimen100=\maxdimen\xdef\r@p@sbbllx{\number\dimen100}
					    \xdef\r@p@sbblly{\number\dimen100}
			                    \xdef\r@p@sbburx{-\number\dimen100}
					    \xdef\r@p@sbbury{-\number\dimen100}}
                        \def\minmaxtest{
			   \ifnum\number\p@intvaluex<\r@p@sbbllx
			      \xdef\r@p@sbbllx{\number\p@intvaluex}\fi
			   \ifnum\number\p@intvaluex>\r@p@sbburx
			      \xdef\r@p@sbburx{\number\p@intvaluex}\fi
			   \ifnum\number\p@intvaluey<\r@p@sbblly
			      \xdef\r@p@sbblly{\number\p@intvaluey}\fi
			   \ifnum\number\p@intvaluey>\r@p@sbbury
			      \xdef\r@p@sbbury{\number\p@intvaluey}\fi
			   }
			\rotate@{\@p@sbbllx}{\@p@sbblly}
			\minmaxtest
			\rotate@{\@p@sbbllx}{\@p@sbbury}
			\minmaxtest
			\rotate@{\@p@sbburx}{\@p@sbblly}
			\minmaxtest
			\rotate@{\@p@sbburx}{\@p@sbbury}
			\minmaxtest
			\edef\@p@sbbllx{\r@p@sbbllx}\edef\@p@sbblly{\r@p@sbblly}
			\edef\@p@sbburx{\r@p@sbburx}\edef\@p@sbbury{\r@p@sbbury}
		\fi
		\count203=\@p@sbburx
		\count204=\@p@sbbury
		\advance\count203 by -\@p@sbbllx
		\advance\count204 by -\@p@sbblly
		\edef\@bbw{\number\count203}
		\edef\@bbh{\number\count204}
}
\def\in@hundreds#1#2#3{\count240=#2 \count241=#3
		     \count100=\count240	
		     \divide\count100 by \count241
		     \count101=\count100
		     \multiply\count101 by \count241
		     \advance\count240 by -\count101
		     \multiply\count240 by 10
		     \count101=\count240	
		     \divide\count101 by \count241
		     \count102=\count101
		     \multiply\count102 by \count241
		     \advance\count240 by -\count102
		     \multiply\count240 by 10
		     \count102=\count240	
		     \divide\count102 by \count241
		     \count200=#1\count205=0
		     \count201=\count200
			\multiply\count201 by \count100
		 	\advance\count205 by \count201
		     \count201=\count200
			\divide\count201 by 10
			\multiply\count201 by \count101
			\advance\count205 by \count201
		     \count201=\count200
			\divide\count201 by 100
			\multiply\count201 by \count102
			\advance\count205 by \count201
		     \edef\@result{\number\count205}
}
\def\compute@wfromh{
		\in@hundreds{\@p@sheight}{\@bbw}{\@bbh}
		\edef\@p@swidth{\@result}
}
\def\compute@hfromw{
	        \in@hundreds{\@p@swidth}{\@bbh}{\@bbw}
		\edef\@p@sheight{\@result}
}
\def\compute@handw{
		\if@height 
			\if@width
			\else
				\compute@wfromh
			\fi
		\else 
			\if@width
				\compute@hfromw
			\else
				\edef\@p@sheight{\@bbh}
				\edef\@p@swidth{\@bbw}
			\fi
		\fi
}
\def\compute@resv{
		\if@rheight \else \edef\@p@srheight{\@p@sheight} \fi
		\if@rwidth \else \edef\@p@srwidth{\@p@swidth} \fi
}
\def\compute@sizes{
	\compute@bb
	\if@scalefirst\if@angle
	\if@width
	   \in@hundreds{\@p@swidth}{\@bbw}{\ps@bbw}
	   \edef\@p@swidth{\@result}
	\fi
	\if@height
	   \in@hundreds{\@p@sheight}{\@bbh}{\ps@bbh}
	   \edef\@p@sheight{\@result}
	\fi
	\fi\fi
	\compute@handw
	\compute@resv}

\def\psfig#1{\vbox {
	\ps@init@parms
	\parse@ps@parms{#1}
	\compute@sizes
	\ifnum\@p@scost<\@psdraft{
		\special{ps::[begin] 	\@p@swidth \space \@p@sheight \space
				\@p@sbbllx \space \@p@sbblly \space
				\@p@sbburx \space \@p@sbbury \space
				startTexFig \space }
		\if@angle
			\special {ps:: \@p@sangle \space rotate \space} 
		\fi
		\if@clip{
			\if@verbose{
				\ps@typeout{(clip)}
			}\fi
			\special{ps:: doclip \space }
		}\fi
		\if@prologfile
		    \special{ps: plotfile \@prologfileval \space } \fi
		\if@decmpr{
			\if@verbose{
				\ps@typeout{psfig: including \@p@sfile.Z \space }
			}\fi
			\special{ps: plotfile "`zcat \@p@sfile.Z" \space }
		}\else{
			\if@verbose{
				\ps@typeout{psfig: including \@p@sfile \space }
			}\fi
			\special{ps: plotfile \@p@sfile \space }
		}\fi
		\if@postlogfile
		    \special{ps: plotfile \@postlogfileval \space } \fi
		\special{ps::[end] endTexFig \space }
		\vbox to \@p@srheight true sp{
			\hbox to \@p@srwidth true sp{
				\hss
			}
		\vss
		}
	}\else{
		\if@draftbox{		
			\hbox{\frame{\vbox to \@p@srheight true sp{
			\vss
			\hbox to \@p@srwidth true sp{ \hss \@p@sfile \hss }
			\vss
			}}}
		}\else{
			\vbox to \@p@srheight true sp{
			\vss
			\hbox to \@p@srwidth true sp{\hss}
			\vss
			}
		}\fi

	}\fi
}}
\psfigRestoreAt

\def\footnoterule{\vfill \vskip -5pt \vglue 6pt \hrule 
width 2in \vglue 6pt}  

\begin{document}
\pagestyle{empty}

\null
\vskip -1.5 truecm
{\baselineskip=22truept\parindent=0pt
{\vbox{\vskip 1.0 truecm\rightskip=0pt plus1fill
{\Large\bf Baryon number segregation at the end of \break
the cosmological quark--hadron transition}
\leftskip=0pt plus1fill }} }

\bigskip
\bigskip
\bigskip
\medskip
\centerline{{\bf Luciano Rezzolla}} \par
\medskip
\bigskip
\medskip
\centerline{{\it International School for Advanced Studies -- SISSA}} 
\medskip
\centerline{Via Beirut 2--4, 34014 Trieste Italy}
\bigskip
\centerline{{\it Physics Department, University of Illinois at 
Urbana-Champaign}}
\medskip
\centerline{1110 West Green St. Urbana IL 61801\footnote{Also: 
NCSA at the University of Illinois at  Urbana-Champaign, Urbana IL 61801}}
\par
\bigskip
\bigskip
\centerline{\tenrm Electronic address:
{\tt rezzolla@astro.physics.uiuc.edu} }\par

\bigskip 
\medskip 
\begin{abstract} 
\bigskip 
\noindent 
One of the most interesting questions regarding a possible first order
cosmological quark--hadron phase transition concerns the final fate of the
baryon number contained within the disconnected quark regions at the end
of the transition. We here present a detailed investigation of the
hydrodynamical evolution of an evaporating quark drop, using a
multi-component fluid description to follow the mechanisms of baryon
number segregation. With this approach, we are able to take account of the
simultaneous effects of baryon number flux suppression at the phase
interface, entropy extraction by means of particles having long
mean-free-paths, and baryon number diffusion. A range of computations has
been performed to investigate the permitted parameter-space and this has
shown that significant baryon number concentrations, perhaps even up to
densities above that of nuclear matter, represent an inevitable outcome
within this scenario. 
\end{abstract}
\bigskip 
\noindent 
PACS number(s): 98.80.Cq, 12.38.Aw, 11.30.Fs \par 
\bigskip 
\centerline {SISSA Ref. 144/96/A} 
\bigskip 
 
\vfill\eject

\setcounter{page}{1}
\pagestyle{plain}
\vsize=20.0truecm 
\hsize=15.0truecm 
\voffset=1.0truecm 
\hoffset=-0.75truecm
\baselineskip=18pt plus 2pt minus 1pt
\parindent=1.0truecm

{\who=1\bigskip\bigskip\medskip\goodbreak
{\Large\bf\noindent\hbox{I.}\hskip 0.5truecm Introduction}\nobreak
\bigskip\medskip\nobreak\who=0}

	If the cosmological quark--hadron phase transition was of first
order with non-negligible supercooling (as we will be assuming in this
paper), then it almost certainly produced inhomogeneities of some kind in
the spatial distribution of baryon number. These inhomogeneities could
have had a relevant influence on the subsequent nucleosynthesis if they
were produced on large enough scales, had sufficiently large amplitude and
contained a significant fraction of the baryon number present in the
Universe \cite{w84}--\cite{ikksl94a}.
	
	Many authors in the past years have calculated the possible
consequences of baryon number inhomogeneities for the subsequent evolution
of the Universe, focussing particularly on the scenarios for inhomogeneous
nucleosynthesis \cite{mm93,jfmk94}. However, in all of these studies it was
necessary to introduce suitably chosen parameters in order to compensate
for lack of knowledge about the spatial distribution and amplitude of the
baryon number peaks which might have been left behind at the end of the
quark--hadron transition. There are various origins for this lack of
knowledge which we will now discuss briefly. A first and major unresolved
issue concerns the typical distance between nucleation sites at the
beginning of the transition. This length scale is not only relevant at the
time when hadron bubbles are nucleated within the slightly supercooled
quark medium, but also represents a characteristic length scale for
subsequent stages of the transition. It is the scale at which neighbouring
hadron bubbles collide and percolate and it also determines the typical
distance between the centres of the disconnected quark regions produced by
bubble coalescence. As a consequence of this, it also represents the
maximum length-scale for the production of peaks in the baryon number
density at the end of the transition. 

	Independently of the scale over which baryon number fluctuations
are produced, the study of baryon number segregation is complicated by the
presence of various mechanisms which could contribute to the scale of the
fluctuations. When there is chemical equilibrium between the two phases,
baryon number density is already ``naturally'' higher in the quark phase
than in the hadron phase. This difference can be further increased as a
result of various other processes which probably act together. One of
these is suppression of baryon number flow across the phase interface
\cite{ah85,fma88,ks88}. Simple statistical considerations suggest that
baryon number cannot be carried entirely together with the hydrodynamical
flow when this moves from the quark to the hadron phase. A
phenomenological explanation for this can be found in the fact that it is
generally more difficult to find in a volume of 1 fm$^3$ and in a time of
$10^{-23}$ sec the right triplet of up and down quarks necessary to form a
color singlet nucleon (either a baryon or an antibaryon) than it is to
find the doublet of quarks necessary to form the lightest hadrons (e.g.
pions). 

	Another mechanism, which could affect the variations in baryon
number density and which is effective during the final stages of
the transition, is long--range energy and momentum transfer by means of
particles having long mean free path (which can be viewed as a form of
radiative transfer). This process, which has recently been investigated by
Rezzolla and Miller \cite{rm96}, involves the relativistic fluids of
strongly interacting particles (i.e. the hadron fluid and the quark
fluid which we will refer to as ``standard fluids'') and the
relativistic fluid of particles having only electromagnetic and weak
interactions (which we refer to as the ``radiation fluid'') with
entropy being extracted from the quark phase by the long
mean-free-path particles of the radiation fluid. A necessary condition
for this process to be significant is that the dimension $R_s$ of the
hadron phase (in the case of bubble growth) or of the quark phase (in
the case of drop evaporation) needs to be roughly comparable with the
smallest mean free path $\lambda$ of the radiation fluid
particles\footnote{Note that in general there
		are two of these length scales, referring to the 
		electromagnetic interaction 
		(relevant for photons, $e^{\pm},\; 
		\mu^{\pm}$, etc., for which $\lambda \simeq 10^4$ fm) 
		and to the weak interaction (relevant for neutrinos and
		antineutrinos, for which $\lambda \simeq 10^{13}$ fm).}.
When this condition is not satisfied, it is a good approximation to
consider the two fluids as either being completely coupled (for $R_s \gg
\lambda$) or completely decoupled (for $R_s \ll \lambda$) and the energy
and momentum interchange between the standard fluid and the radiation
fluid is then either maximally efficient or nonexistent. 

	Besides the baryon number segregating mechanisms mentioned above,
there is also another competing process which needs to be taken into
account. The formation of localized regions of high baryon number density
is counteracted by baryon number diffusion which occurs whenever a
deviation from homogeneity is produced and is effective both during the
quark--hadron transition and after it. 

	While, from the above discussion, the formation of peaks in baryon
number density appears to be a rather inevitable consequence of a first
order quark--hadron transition, a number of quantitative aspects of this
picture still remain to be fully investigated. For this reason, and before
the microphysics of baryon number flow across the phase interface or of
baryon number diffusion is known in more detail, we here present results
from systematic numerical computations of baryon number concentration
during the evaporation of an isolated quark drop. These simulations are an
extension of earlier calculations of the final stages of the quark--hadron
transition, in which the hydrodynamics of a spherical isolated quark drop
was coupled to the processes of relativistic radiative transfer between
the standard fluids and the radiation fluid. In the present simulations,
we have also studied the behaviour of the additional distinct ``fluid'' of
baryon number carriers and have followed the progress of the segregation
as it is promoted by the suppression mechanism at the interface and by the
entropy extraction but moderated by baryon diffusion.
	
	Results from computations performed with different values of the
relevant parameters are presented in Section V. Before this, Section II
contains a brief review of the mechanisms responsible for baryon number
segregation and of their role during the various stages of the transition.
Section III presents the system of relativistic equations which we use for
calculating the hydrodynamics of the quark drop and the radiative transfer
processes. (The formal derivation of these equations is not given here but
can be found in references \cite{rm94,mr95,rmp95,rm96}.) In Section IV we
derive a relativistic equation for the diffusion of baryon number and
write it in a convenient finite difference form. This Section also
introduces a phenomenological analytical expression by means of which the
suppression mechanisms at the interface can be represented.  Section VI
contains the conclusions and also a discussion of our results in relation
to the scenarios for inhomogeneous cosmological nucleosynthesis and for
the production of primordial magnetic fields. We adopt units for which $c
= \hbar = k_{_B} = 1$ and use commas or the usual ``$\partial$'' notation
to denote partial derivatives. 

{\who=1\bigskip\bigskip\medskip\goodbreak
{\Large\bf\noindent
\hbox{II.}\hskip 0.5truecm Evolution of the baryon number concentration}
\nobreak \bigskip\medskip\nobreak\who=0}

	As mentioned in the Introduction, production of inhomogeneities in
baryon number density can be affected by a number of different factors. In
this Section we outline a schematic picture of the various stages of the
transition during which baryon number segregation can either be favoured
or impeded. First, we note that even in the absence of specific mechanisms
for this segregation, there is a ``natural'' tendency for the baryon
number density to be higher in the quark phase than in the hadron phase
since baryon number is carried by almost massless quarks in the quark
phase, whereas in the hadron phase it is carried by heavy nucleons whose
number density is suppressed by an exponential factor. An estimate for the
contrast in baryon number density resulting from this can easily be
calculated for stages at which global chemical equilibrium is near to
holding. The chemical potentials in the two phases can then be set equal
($\mu^q_b=\mu^h_b=\mu_b$) signifying that, if the phase interface were not
in motion, equal fluxes of baryon number would cross it in both
directions.
	
	Because of the complexity of QCD and of the spectrum of hadronic
species present in the low temperature (hadron) phase, it is necessary to
make approximations if we want to obtain simple analytical expressions for
the baryon number density in the two phases. Following the approach of our
earlier papers, we use the phenomenological bag model for the quark phase
(treating the quark-gluon plasma as consisting of relativistic
non-interacting particles within a false vacuum) and the energy density
and pressure may then be written as

\begin{equation}
\label{eosq}
e_q = g^*_q \left({\pi^2 \over {30}}\right) T^4_q + B \ ,
\hskip 1.0truecm
p_q = g^*_q \left({\pi^2 \over {90}}\right) T^4_q - B \ ,
\end{equation}

\medskip 
\noindent 
where $g^*_q$ is the effective number of degrees of freedom in the quark
phase (including also the relevant number of degrees of freedom for the
particles of the radiation fluid when the radiation and standard fluids
are fully coupled \cite{rm96}). The baryon number density is then given by

\begin{equation}
\label{bndq}
n^q_b \simeq {N_c N_f \over {27}} 
\left ( {\mu_b \over {T}} \right) T^3 , 
 \end{equation}

\noindent
where $N_c$ and $N_f$ are, respectively, the number of colours (three) and
the number of quark flavours (which we will take here to be two). On a
similar level of approximation, we can consider the pion contribution to
the grand partition function of the hadron phase as being the dominant one
coming from the whole spectrum of hadron species, neglect finite volume
effects and the repulsive interaction between hadrons and describe the
hadron phase as being essentially a gas of relativistic massless particles 
with the equation of state

\begin{equation}
\label{eosh}
e_h = g^*_h \left({\pi^2 \over {30}}\right) T^4_h \ ,
\hskip 4.0truecm p_h = {1 \over 3} e_h \ ,
\end{equation}

\medskip
\noindent
 where $g^*_h$ is the effective number of degrees of freedom in the hadron
phase (equivalent to $g^*_q$ in the quark phase). It is then possible to
obtain a simple and compact expression for the baryon number density
in the hadron phase as 

\begin{eqnarray}
\label{bndh}
n^h_b =  \left ( {2 m T \over {\pi}} \right)^{3/2} 
{\rm sinh} \left ( {\mu_b \over T} \right)
e^{-m/T} 
\simeq  \left ( {2 m T \over {\pi}} \right)^{3/2} 
\left ( {\mu_b \over T} \right) 
e^{-m/T} \;,
\end{eqnarray}

\noindent
 with $m$ being the nucleon mass.

	With all of these assumptions, the contrast in baryon number
density between the two phases is given by

\begin{equation}
\label{bndcont}
R={n^q_b\over n^h_b}= {N_f \over 9} 
\left ( {\pi T \over {2 m}} \right)^{3/2}
e^{m/T} \sim 10 \ \ \ {\rm for}\ T=T_c=150\ {\rm MeV}, 
\end{equation}

\noindent
where $T_c$ is the critical temperature for the transition. Similar
estimates of the baryon number contrast have been found also from more
detailed analyses of the baryon number density in the two phases
\cite{gpga95}.

	It is important to stress that when the hypothesis of global
chemical equilibrium is valid, equation (\ref{bndcont}) holds between any
two generic points in the two phases. It is reasonable to assume that
global chemical equilibrium holds when the velocity of the phase interface
is much smaller than the diffusion velocity of baryon number and the
length scales of the disconnected hadron or quark regions are much smaller
than the baryon number diffusion length-scale during the transition. This
is the case, for example, soon after bubble nucleation, when most of the
bubbles present have radii around the critical value

\begin{equation}
\label{cr}
R_c= {2 \sigma \over {p_h-p_q}} \;,
\end{equation}

\noindent
where $\sigma$ is the surface tension associated with the phase interface. 

	After nucleation, the bubble starts to expand (the low temperature
phase is thermodynamically favoured) and its surface accelerates until it
reaches a steady state velocity corresponding to self similar growth for a
weak deflagration bubble \cite{mp89}. (We are working here within the
scenario of small supercooling, in which weak deflagrations are the
relevant mechanism for the motion of the phase interface, and will assume
that the separation between the nucleation sites is not very small.) The
expansion produces a deviation away from chemical equilibrium and the
baryon number density can be increased in the vicinity of the interface
\cite{ks88}, piling up mostly on the quark side but also to some extent in
the hadron phase as a result of the increased flux coming from the quark
medium. The situation in the bulk of the two phases is not significantly
modified, however. Diffusion tends to impede the accumulation of baryon
number and, when the {\it self similar} growth stage is reached for the
hydrodynamic variables, the baryon number density attains a {\it
stationary} profile. (Note that the intrinsic length scale set by the
diffusion coefficient prevents the baryon number density following a
self similar solution.) Most of the excess baryon number on the quark side
of the interface is then within a layer of thickness $r_d$ with the value
of the baryon number density there being joined to the background value in
the bulk of the quark phase via a decaying profile. A rough estimate for
$r_d$ is given by

\begin{equation}
\label{width}
r_d \approx {D\over {v_f}} \;,
\end{equation}

\noindent
where $D$ is the baryon number diffusion coefficient and $v_f$ is the
steady state velocity of the front as measured from the centre of the
bubble.

	The self similar growth is first broken by the coupling
between the radiation fluid and the standard fluids when the bubble
has reached dimensions of the order of $\lambda$ \cite{mr95}. After
the coupling however, the similarity solution for bubble growth is
restored and it ultimately ceases when the spherical compression waves
preceding the deflagration fronts of neighbouring bubbles start to
interact. (This occurs well before the bubble surfaces themselves come
into contact.)  In principle, for a perfect fluid, the compression
waves would be fronted by shocks but it should be noted that the
predicted amplitude of such shocks is negligibly small for the case of
spherical bubbles and small supercooling which we are considering here
\cite{mp90,ikksl94b,mr95,ksl95b}. However, independently of whether there
is any significant shock or not, the kinetic energy of the ordered
motion will be progressively converted into internal energy via
compression, once adjacent bubbles have started to interact, and the
temperatures of both phases will be raised near to the critical
temperature $T_c$. At this stage, the bubbles grow much more slowly,
on a time scale which is essentially set by the expansion of the
Universe leading to cooling which allows for continuation of the
transition. In this slow growth stage, global chemical equilibrium is
restored and the baryon number which has accumulated near to the phase
interface can be redistributed into the bulk of the two phases. When
the surfaces of adjacent bubbles meet, they coalesce to form larger
bubbles thus minimizing the total surface energy.  This coalescence
gives rise to disconnected quark regions which then proceed to
evaporate, tending to become spherical under the action of surface
tension.

	The hydrodynamical evolution of isolated evaporating quark drops
has been investigated in detail in a number of studies
\cite{kks86,rmp95,rm96,ksl95b} and can be essentially summarized as
consisting of a self similar stage followed by one in which long range
energy and momentum transfer takes place and then a final decay which may
be dominated by surface tension effects and become increasingly rapid. 

	During the first stage, quark drops evaporate converting quarks
into hadrons at the rate necessary to keep the internal compression
constant and uniform. In a very pictorial description, each quark drop can
be viewed as behaving like a shrinking ``leaky balloon'' which is ejecting
material at the rate necessary to avoid producing any compression. If
baryon number were entirely carried along with the hydrodynamical flow and
no suppression of baryon number flux occurred at the interface, then this
stage of the evaporation would not produce any increases in the baryon
number density. If, however, there {\it is} some flux suppression, this
stage will still consist of a self similar evolution for all of the
hydrodynamical variables apart from the baryon number which will
accumulate ahead of the interface with diffusion to either side of it in a
way similar to that discussed above for a growing hadron bubble. 

	When the quark drop reaches dimensions comparable with the mean
free path for the particles of the radiation fluid, the process of
entropy extraction breaks the self similarity of the evaporation and
produces a sharp increase in the fluid compression (of both the quark
and hadron phases) \cite{rm96} and also in the baryon number density.
It should be stressed that this increase in baryon number density is
distinct from the one produced by flux suppression and is not
necessarily localized at the phase interface but is rather extended
over the whole quark region. This is because its origin is not
hydrodynamical but is the consequence of the long range energy and
momentum transfer via the radiation fluid particles.

	As pointed out in \cite{rmp95}, the very final stages of the drop
evaporation (for $R_s {\hbox {$\;\raise.4ex\hbox
{$<$}\kern-.75em\lower.7ex\hbox{$\sim$} \;$}} 10^2$ fm) may be dominated
by the surface tension $\sigma$ which could give rise to an increasingly
rapid evaporation and a consequent increase of both the compression and
the baryon number density. However, recent lattice gauge calculations seem
to indicate rather small values for the surface tension coefficient
$\sigma_0=\sigma/T^3_c\approx 0.01-0.02$ \cite{ikksl94b} and, if this is
correct, then surface tension would play only a minor role for the
increase of baryon number density. 

	In the next Section, we introduce the set of relativistic
hydrodynamical equations which we have used for describing the dynamics
of single isolated evaporating quark drops and from whose numerical
solution we have derived quantitative results for the profile and
amplitude of the baryon number inhomogeneity produced at the end of the
quark--hadron transition.

{\who=1\bigskip\bigskip\medskip\goodbreak
{\Large\bf\noindent
\hbox{III.}\hskip 0.5truecm Relativistic hydrodynamical equations}
\nobreak \bigskip\medskip\nobreak\who=0}

	We here briefly review the relativistic equations which we have
used for calculating baryon number concentration at the end of the
quark--hadron transition and refer the reader to previous papers
\cite{rm94,mr95,rmp95,rm96} for further details. The equations can be
divided into three main groups, related to: {\it i)} the dynamics of the
spherical drop, {\it ii)} radiative transfer, {\it iii)} the behaviour of
the diffusing ``baryon number fluid''. We use a Lagrangian spherically
symmetric coordinate system comoving with the fluid and having its origin
at the centre of the drop. It is appropriate to write the metric line
element as

\begin{equation}
\label{metric}
ds^2 = -a^2 dt^2 + b^2 d\mu^2 +
R^2 ( d\theta^2 + {\rm sin}^2 \theta \ d\phi^2) ,
\end{equation}

\medskip 
\noindent 
where $\mu$ is the comoving radial coordinate and $R$ is the associated
Eulerian coordinate. Local conservation of energy and momentum for the
combined fluids (the standard fluid together with the radiation fluid), as
measured in the local rest frames of the standard fluid, can be expressed 
as the vanishing of the projections of the combined stress-energy tensor
parallel and perpendicular to the direction of the fluid four-velocity
{\bf u}. Using this, the continuity equation and the Einstein equations,
the full set of equations for determining the behaviour of the standard
fluids can be written as: 

\begin{equation}
\label{radialmom}
u_{,\> t}=-a\biggl[{{\Gamma\over b}\biggl({p_{,\>\mu}+
bs_1\over {e+p}}\biggr)
+ 4\pi G R \biggl({p+{1\over 3}w_0 + w_2}\biggr)
+ {G M\over {R^2}}}\biggr] ,
\end{equation}

\begin{equation}
\label{energy}
e_{,\> t}=w\rho_{,\> t}-as_0 ,
\end{equation}

\begin{equation}
\label{contin}
{(\rho R^2)_{,\> t}\over {\rho R^2}}=
-a\biggl({{u_{,\>\mu}-4\pi G b R w_1\over {R_{,\>\mu}}}}\biggr) ,
\end{equation}

\begin{equation}
\label{constr}
{(aw)_{,\>\mu}\over {aw}}=
-{w\rho_{,\>\mu}-e_{,\>\mu}+bs_1\over {\rho w}} ,
\end{equation}

\begin{equation}
\label{emmu}
M_{,\>\mu}=4\pi R^2 R_{,\>\mu}\biggl({e + w_0 +
{u\over {\Gamma}}w_1}\biggr) ,
\end{equation}

\begin{equation}
\label{b}
b = {1 \over { 4 \pi R^2 \rho }}\ \;,
\hskip 2.0 truecm
u={1\over a}R_{,\> t} \;,
\end{equation}

\begin{equation}
\label{gamma}
\Gamma={1\over b}R_{,\>\mu}=
\biggl({1+u^2-{2GM\over R}}\biggr)^{\!1/2} ,
\end{equation}

\medskip 
\noindent 
 where $\ w=(e+p)/\rho\ $ is the specific enthalpy of the standard fluids,
$u$ is the radial component of the fluid four velocity in the associated
Eulerian frame, $\Gamma$ is the general relativistic analogue of the
Lorentz factor and $M$ a generalized mass function. The quantity $\rho$ is
the compression factor which expresses the variation in proper volume of
comoving elements of the standard fluids with respect to a fiducial value.
(For a classical standard fluid, $\rho$ can be taken to represent the rest
mass density). The interaction between the standard fluids and the
radiation fluid enters through the source functions $s_0$ and $s_1$ and
through the radiation contributions to the gravitational terms via the
radiation energy density $w_0$, energy flux $w_1$ and anisotropy $w_2$.
The latter quantities are calculated by considering local conservation of
energy and momentum of the radiation fluid alone, and this provides the
following set of equations \cite{rm94}

\begin{equation}
\label{w0}
(w_0)_{,\> t}+{a\over b}(w_1)_{,\>\mu}
+{4\over 3}\biggl({{b_{,\> t}\over b}+{2R_{,\> t}\over R}}\biggr)w_0
+{2a\over b}\biggl({{a_{,\>\mu}\over a}+{R_{,\>\mu}\over R}}\biggr)w_1
+\biggl({{b_{,\> t}\over b}-{R_{,\> t}\over R}}\biggr)w_2=as_0 ,
\end{equation}

\begin{equation}
\label{w1}
(w_1)_{,\> t}+{a\over b}\biggl({{1\over 3}w_0+w_2}\biggr)_{\!\! ,\mu} +
{4a_{,\>\mu}\over {3b}}w_0
+2\biggl({{b_{,\> t}\over b}+{R_{,\> t}\over R}}\biggr)w_1
+{a\over b}\biggl({ {a_{,\>\mu}\over a}+{3R_{,\>\mu}\over R}}\biggr)w_2
=as_1 ,
\end{equation}

\begin{equation}
\label{clrl}
w_2 = f_{\!_E} w_0 \  ,
\end{equation}

\noindent
\medskip
where (\ref{clrl}) is a closure relation written in terms of a variable
Eddington factor $f_{\!_E}$. In our calculations, the source functions
$s_0$, $s_1$ are expressed as

\begin{equation}
\label{s0_s1}
s_0 = {(1+\alpha_2) \over {\lambda}} 
\left[ g_r \left({\pi^2 \over{30}}\right) T_{_F}^4  - w_0\right]\ \;,
\hskip 2.0 truecm 
s_1 = - {w_1 \over {\lambda}} ,
\end{equation}

\medskip 
\noindent 
with $g_r$ and $T_F$ being the number of degrees of freedom and the
temperature of the radiation fluid. The parameter $\alpha_2$ is an
adjustable coefficient (ranging between zero and one) which takes account
of the contribution of non-conservative scatterings to the energy source
function (for the present calculations we take $\alpha_2=1$). We treat the
phase interface as a discontinuity surface and track it continuously
through the finite difference grid using junction conditions which are
imposed across the interface and a characteristic form of the above
hydrodynamic equations which is used for the flow regions adjacent to it. A
detailed discussion of these extra equations and of the way in which they
are obtained has been given in previous papers \cite{rm94,mr95,rm96} and
we will not repeat it here but, rather, just list the expressions used for
the junction conditions. The energy and momentum junction conditions for
the standard fluids are expressed as

\begin{equation}
\label{esfjc}
[(e+p)ab]^{\pm}=0 ,
\end{equation}

\begin{equation}
\label{msfjc}
[eb^2{\dot \mu_s}^2+pa^2]^{\pm}=-{\sigma f^2\over 2}\biggl\{
{{1\over {ab}}{d\over {dt}}\biggl({b^2\dot \mu_s\over f}\biggr) +
{f_{,\>\mu}\over {ab}} +
{2\over {fR}}{(b\dot \mu_s u + a\Gamma)}}\biggr\}^{\pm} .
\end{equation}

\medskip
\noindent
As junction conditions for the energy and momentum of the radiation fluid
after the decoupling from the standard fluids has begun to take place 
\cite{rm96} we use: 

\begin{equation}
\label{erfjc}
\biggl[{ab\dot \mu_s\biggl({{4\over 3}+f_{\!_E}}\biggr)
w_0-(a^2+b^2{\dot \mu_s}^2)w_1}\biggr]^{\pm}=0 ,
\end{equation}

\begin{equation}
\label{msrjc}
\biggl[{\left\{{a^2\biggl({{1\over 3} + f_{\!_E}}\biggr)+
b^2{\dot \mu_s}^2}
\right\}w_0-2ab\dot \mu_s w_1}\biggr]^{\pm}=0\ ,
\end{equation}

\medskip
\noindent
while for the metric coefficients we have:

\begin{equation}
[R]^{\pm}=0\ ,
\end{equation}

\begin{equation}
[au+b \dot \mu_s \Gamma]^{\pm}=0\ ,
\end{equation}

\begin{equation}
\label{mudot}
[a^2-b^2 {\dot \mu_s}^2]^{\pm}=0\ .
\end{equation}  

\medskip
\noindent
Note that the use of brackets in equations (\ref{esfjc})--(\ref{mudot})
follows the conventions: \hbox{$[A]^{\pm} =$} \break $A^+ - A^-$,
\hbox{$\{A\}^{\pm} = A^+ + A^-$}, and that the superscripts $^{\pm}$
indicate quantities immediately ahead of and behind the interface. The
interface location is denoted by $\mu_s$ with \hbox{$\dot \mu_s =
d\mu_s/dt$} and \hbox{$f =(a^2 - b^2 \dot \mu^2_{_S} )^{1/2}$}. 

	Finally, for a phase interface which behaves as a weak
deflagration front, it is necessary to supply one further equation giving
the rate at which material is converted from one phase to the other (or,
alternatively, to specify the interface velocity). As discussed in
previous papers (see, for example \cite{mp89}) we use a formula giving the
rate of conversion of quarks into hadrons in terms of a black-body model
for the phase interface, suitably corrected with an adjustable
accommodation coefficient $\alpha_1$ ($0 \leq \alpha_1 \leq 1$) to take
account of deviations away from an ideal black-body law. We then have

\begin{eqnarray}
\label{flux}
-{{aw\dot\mu_s}
\over{4\pi R^2_{_S}(a^2-b^2\dot \mu_s^2)}}
=\left({\alpha_1 \over 4}\right)
(g_h+g_r) \left({\pi^2 \over {30}}\right)
(T_q^4-T_h^4)\ .
\end{eqnarray}

\noindent
 For the calculations presented here, we have taken $\alpha_1=1$, but an
extend discussion of computations with smaller values of $\alpha_1=1$ can
be found in \cite{rm96}. 

{\who=1\bigskip\bigskip\medskip\goodbreak
{\Large\bf\noindent\hbox{IV.}\hskip 0.5truecm Baryon number flux and 
diffusion} \nobreak \bigskip\medskip\nobreak\who=0}

	As mentioned in the Introduction, simple considerations seem to
suggest that baryon number would not be carried along entirely together
with the hydrodynamical flow but, rather, that it would tend to be
``trapped'' inside the high temperature phase because of flux
suppression at the phase interface. In order to study the evolution of
the baryon number distribution, we need to extend the multi-component
fluid description discussed in Section III by adding to the treatment
of the standard fluids and the radiation fluid, an additional ``baryon
number fluid''. In particular, we need to introduce equations to
describe the hydrodynamics of a fluid which has a suppressed flow at
the phase interface and which diffuses relative to the standard fluids.

	The effects of this diffusion can be included by introducing a
diffusive flux four-vector {\bf q} into the baryon number continuity
equation which becomes

\begin{equation}
\label{cont_nb}
\left( n_b u^{\alpha} + q^{\alpha} \right)_{;\ \alpha} = 0 \ ,
\end{equation}

\medskip 
\noindent 
 where $n_b$ is the baryon number density, {\bf u} is still the four
velocity of the {\it standard} fluid and Greek indices are taken to run
from 0 to 3. In the frame comoving with the standard fluid,
\hbox{$u^{\alpha}\equiv (1/a,\; 0,\; 0,\; 0)$} and {\bf q} has only a
spatial component [i.e. $u_{\alpha}q^{\alpha}=0\;$ and $\;q^{\alpha}\equiv
(0,\; q^{\mu},\; 0,\; 0)$]. Using the property \hbox{$(V^{\alpha})_{;\
\alpha}= (\sqrt{-g} V^{\alpha})_{,\ \alpha}/\sqrt{-g}\;$} (where $g$ is
the determinant of the metric tensor) and the metric relations
(\ref{metric}), (\ref{gamma}), we can rewrite (\ref{cont_nb}) as

\begin{equation}
\label{cont_nb1}
b R^2 \left[{\partial (n_b)\over {\partial t}}  + 
\left({a n_b \over {R^2}}\right) 
{\partial (u R^2) \over {\partial R}} \right] = 
{a R^2 \over \Gamma} {\partial (b \Gamma)\over {\partial R}}
{\partial (D n_b) \over {\partial R}} +
a b^2 \Gamma {\partial \over {\partial R}}
\left[ {R^2 \over  {b \Gamma}} 
{\partial (D n_b) \over {\partial R}} \right] 
\ , \end{equation}

\medskip \noindent 
(see the Appendix for details) where the radial component of the
diffusive flux is written as

\begin{equation}
\label{qmu}
q^{\mu} \equiv - {1\over{(b\Gamma)^2}} 
{\partial (D n_b)\over {\partial \mu}} =
- {1\over{b\Gamma}} {\partial (D n_b)\over{\partial R}}. 
\end{equation}
	
	In the Newtonian limit, $\Gamma=a=1$ and equation (\ref{cont_nb1})
reduces to the standard Lagrangian diffusion equation in spherical
coordinates, with $D$ being the diffusion coefficient (note that here,
$\partial/\partial t$ is a Lagrangian time derivative for a location
comoving with the standard fluid).  An attractive feature of Lagrangian
schemes is that advection is treated exactly. Exploiting this, it is
convenient to use for the finite-difference representation of equation
(\ref{cont_nb1}), the simple and compact expression:

\begin{eqnarray}
\label{fntdif}
4 \pi (n_b)^{n+1}_{j+1/2} 
\left [ {(R^2)^{n+1}_{j+1/2} \over {\Gamma^{n+1}_{j+1/2}} } \right ] 
\Delta R^{n+1}_{j+1/2}
= 4 \pi (n_b)^{n}_{j+1/2} 
\left [ {(R^2)^{n}_{j+1/2} \over {\Gamma^{n}_{j+1/2}} } \right ] 
\Delta R^{n}_{j+1/2} 
\hskip 2.0truecm \nonumber \\ \nonumber \\ \hskip 2.5truecm
-\ {4 \pi\over {\Gamma^{n+1/2}_{j+1/2}}} \left [ (R^2)^{n}_{j+1} 
(\Phi_D)^{n}_{j+1} - (R^2)^{n}_{j} (\Phi_D)^{n}_{j} \right ] 
a^{n+1/2}_{j+1/2} \Delta t^{n+1/2}_{j+1/2}\ , \nonumber \\  
\end{eqnarray}

\medskip\noindent
where the superscripts refer to the time level at which the quantity is
calculated and the subscripts to the position in the spatial grid,
\hbox{ $\Delta R^{n}_{j+1/2}= R^{n}_{j+1}-R^{n}_{j}$ } and

\begin{equation}
\label{phid}
(\Phi_D)^{n}_{j}= -D \left[ { (n_b)^{n}_{j+1/2}-(n_b)^{n}_{j-1/2} 
\over{R^{n}_{j+1/2}-R^{n}_{j-1/2}} } \right]\ ,
\end{equation}

\medskip\noindent
is the diffusive flux of baryon number. A rough estimate for the value of
the diffusion coefficient $D$ (which we take to be constant in time,
uniform in space and the same in both phases) can be deduced with rather
simple arguments if we rule out the possibility of the diffusion being
turbulent \cite{afmm89} (which seems likely to be correct \cite{jfmk94}).
In this case, baryon number diffusion can be described as a simple
Brownian motion of baryon number carriers having a mean free path of the
order \hbox{$\lambda_{free} \;\raise.4ex\hbox{$>$}\kern-
.75em\lower.7ex\hbox{$\sim$} \; T^{-1}_c$} \cite{ks88} giving a
microscopic diffusion coefficient $D \sim 10^{-1}-10$ fm. 

	Clearly, the baryon number within a given grid zone changes in
time only if a baryon flux crosses its boundaries. While for the bulk of
the two phases, the flux leading to the variation is just the diffusive
flux (\ref{phid}), a special treatment is necessary for the two parts of
the grid zone containing the phase interface. At the interface, the
diffusive flux is accompanied by a much larger flux which is related to
the hydrodynamical flow $\Phi_{\rho}$ of elements of the quark gluon
plasma as it is converted to the hadron phase. As mentioned earlier, this
flux could be subject to suppression processes at the interface but no
exact expression is yet available for this. We therefore proceed here by
defining a phenomenological expression for the net baryon number flux
across the interface $\Phi_{b}$ in terms of the hydrodynamical flux
$\Phi_{\rho}$ and a suitable ``filter factor'' $F$ which expresses the
ratio between the baryon number passing across the phase interface and the
total baryon number incident on it. 

	In principle, $F$ could be expressed in terms of the probability
of finding (from all of the quarks and antiquarks present) three quarks of
the right types within a volume of 1 fm$^3$ and in a time of $10^{-23}$ s,
in terms of the ``transparency'' of the phase interface to the passage of
a baryon number carrier from the quark phase to the hadron phase $\Sigma_{q
\rightarrow h}$ and of the corresponding probability $\Sigma_{h
\rightarrow q}$ that a baryon hitting the phase boundary from the hadron
phase is absorbed \cite{fma88}. Moreover, referring to a situation in
which chemical equilibrium holds, it would be possible to express
$\Sigma_{q \rightarrow h}$ in terms of $\Sigma_{h \rightarrow q}$ and this
would restrict the uncertainty to this latter quantity only.  (Although
convenient, this approach requires that the baryon transmission
probability does not vary significantly when the equilibrium is broken.)

	Unfortunately, no reliable value for the baryon transmission
probability $\Sigma_{h \rightarrow q}$ is known at present and, worse than
this, different approaches to the study of the rates of elementary
processes taking place at the phase interface seem to result in quite
different estimates of it (see \cite{fma88,skam90} and \cite{bp93} for
further references). In view of this uncertainty, we will treat the filter
factor as essentially a free parameter, adopting the reference value
$F\sim 10^{-1}$ as estimated from the expressions presented by Fuller et
al. \cite{fma88} for $T_c=150$ MeV and $\Sigma_{h \rightarrow q}\sim
10^{-3}$. (The value $F=1$ corresponds to the case where the baryon number
flow crosses the phase interface unimpeded and clearly represents an upper
limit.)

	The flux of elements of the standard fluid across the interface
can be evaluated by projecting the flux four-vector along the unit
spacelike four-vector {\bf n} normal to the timelike hypersurface
describing the time evolution of the interface. We obtain

\begin{equation}
\label{phirho}
\Phi_{\rho}\equiv \rho u^{\alpha} n_{\alpha} =
- { {\dot \mu_s} \over { 4 \pi R^2_s f} }\ ,
\end{equation}

\medskip\noindent
where ${\dot \mu_s} <0$, $f=(a^2 -b^2 {\dot \mu_s}^2)^{1/2}$ and

\begin{equation}
u^{\alpha} \equiv {1\over a} (1,\ 0,\ 0,\ 0), 
\hskip 2.0truecm 
n_{\alpha} \equiv {ab\over f} (-{\dot \mu_s},\ 1,\ 0,\ 0)\ .
\end{equation}

\medskip\noindent
Neglecting diffusive contributions to the baryon number flux across the
interface, we obtain

\begin{equation}
\label{phib}
\Phi_b=F \left ({n^q_b \over {\rho_q}}\right ) \Phi_{\rho} = 
- F \left ({n^q_b \over {\rho_q}}\right )
\left ({ {\dot \mu_s} \over { 4 \pi R^2_s f} }\right) \ ,
\end{equation}

\medskip\noindent
and we assume continuity of this flux across the interface (i.e. 
$[\Phi_b]^{\pm}=0$).

	In the next Section we will present results from numerical
computations made using the set of equations which we have introduced
here and illustrate the roles played in the segregation of baryon
number by suppression mechanisms and by radiative transfer.

{\who=1\bigskip\bigskip\medskip\goodbreak
{\Large\bf\noindent\hbox{V.}\hskip 0.5truecm Numerical strategy and 
results} \nobreak \bigskip\medskip\nobreak\who=0}

	The numerical approach implemented for the solution of the
present set of equations (\ref{radialmom})--(\ref{gamma}),
(\ref{w0})--(\ref{s0_s1}), (\ref{esfjc})--(\ref{mudot}) and
(\ref{cont_nb1}) is based on that used for our earlier computation of
the evaporation of a quark drop including the effects of long range
energy and momentum transfer \cite{rm96} with an extension to include
the parabolic diffusion equation (\ref{cont_nb1}).

	As is frequently the case for numerical computations in which a
diffusion equation needs to be solved, we are here faced with the
problem of performing calculations with a mixed set of equations and
avoiding excessive restrictions placed on the time step by the von
Neumann criterion for stability in solving the parabolic equation.
Obeying this criterion, which is more stringent than the usual Courant
one since it depends quadratically on the minimum grid spacing and is
inversely proportional to the diffusion coefficient \cite{p80}, allows
one to perform an explicit integration of equation (\ref{fntdif})
without any further constraint. However, for the present case,
implementing the von Neumann condition produced computational times
for each simulation which were not affordable. The reason for this is
connected with the particular organization of our grid which has an
exponentially increasing spacing in order to facilitate following the
dynamics of the drop through several orders of magnitude change in the
radius. To avoid this problem, we have implemented a standard
``flux limiter'' scheme in which a control is set on the diffusive flux
in the quark phase preventing it from evacuating the baryon number
content of any grid zone within a single time-step \cite{ob87}. The
results obtained in this way were found to be in excellent agreement
with ones from a comparison calculation without the flux limiter and
using the von Neumann condition (but which required a computational
time longer by a factor of twenty).

	Another concern in the present calculations has been that of
preserving as closely as possible the overall conservation of baryon
number.  Clearly results for the final distribution of baryon number at
the end of the transition will be worthless if the integration of the
diffusion equation (\ref{fntdif}) is not accurate enough and is
significantly producing or destroying baryon number. This equation is
nearly conservative in form but special attention needed to be paid to
the calculation of interpolated values for $a$ and $\Gamma$ (for which
we used function fitting procedures) and to the implementation of the
regridding procedure \cite{rmp95,rm96}. Having done this, our
computations preserve the total baryon number in the grid to an
accuracy of a few parts in $10^8$ for runs of about $10^6$ time steps.

	In the following, we present results from computations which have
been performed evolving from initial data given by the self similar
solutions derived in \cite{rmp95} for a spherical quark drop with
initial dimensions $R_{s,0}=10^7$ fm, at an initial temperature ${\hat
T_q} = T_q/T_c = 0.998$, surrounded by a hadron plasma at temperature
${\hat T_h} = T_h/T_c = 0.990$. The similarity solutions provide
suitable initial conditions for all of the variables apart from the
baryon number which does not necessarily follow the bulk hydrodynamical
flow. We decided to start with the baryon number density in each phase
being uniform and given by the expressions (\ref{bndq}) and
(\ref{bndh}) presented in Section II which correspond to conditions of
chemical equilibrium. This is very approximate because global chemical
equilibrium does not apply for a situation with a moving interface such
as the one which we are considering but, nevertheless, initial
conditions imposed in this way are sufficiently good to allow the
solution to relax rapidly to a consistent one.

	We have examined the effects on the overall solution of varying the
values of important input parameters and we will be discussing this in
detail but first we concentrate on the results obtained for a set of
fiducial parameter values. This ``standard'' run follows the
evaporation of a quark drop with surface tension parameter $\sigma_0 =
0.01$, filter factor $F=0.3$ and diffusion coefficient $D=1$ fm. The
decoupling radius $R_d$ (i.e. the drop radius at which the decoupling
between the standard fluids and the radiation fluid is allowed to take
place \cite{rm96}) is set to be $10^4$ fm, which corresponds to the
average mean free path of the electromagnetically-interacting
particles. We do not consider here a decoupling with neutrinos which,
however, would follow a similar hydrodynamical behaviour except for the
different number of degrees of freedom involved \cite{rm96}. 

\vfill\eject

\vbox{ \vskip -1.75truecm \hskip -1.75truecm
\centerline{
{\psfig{figure=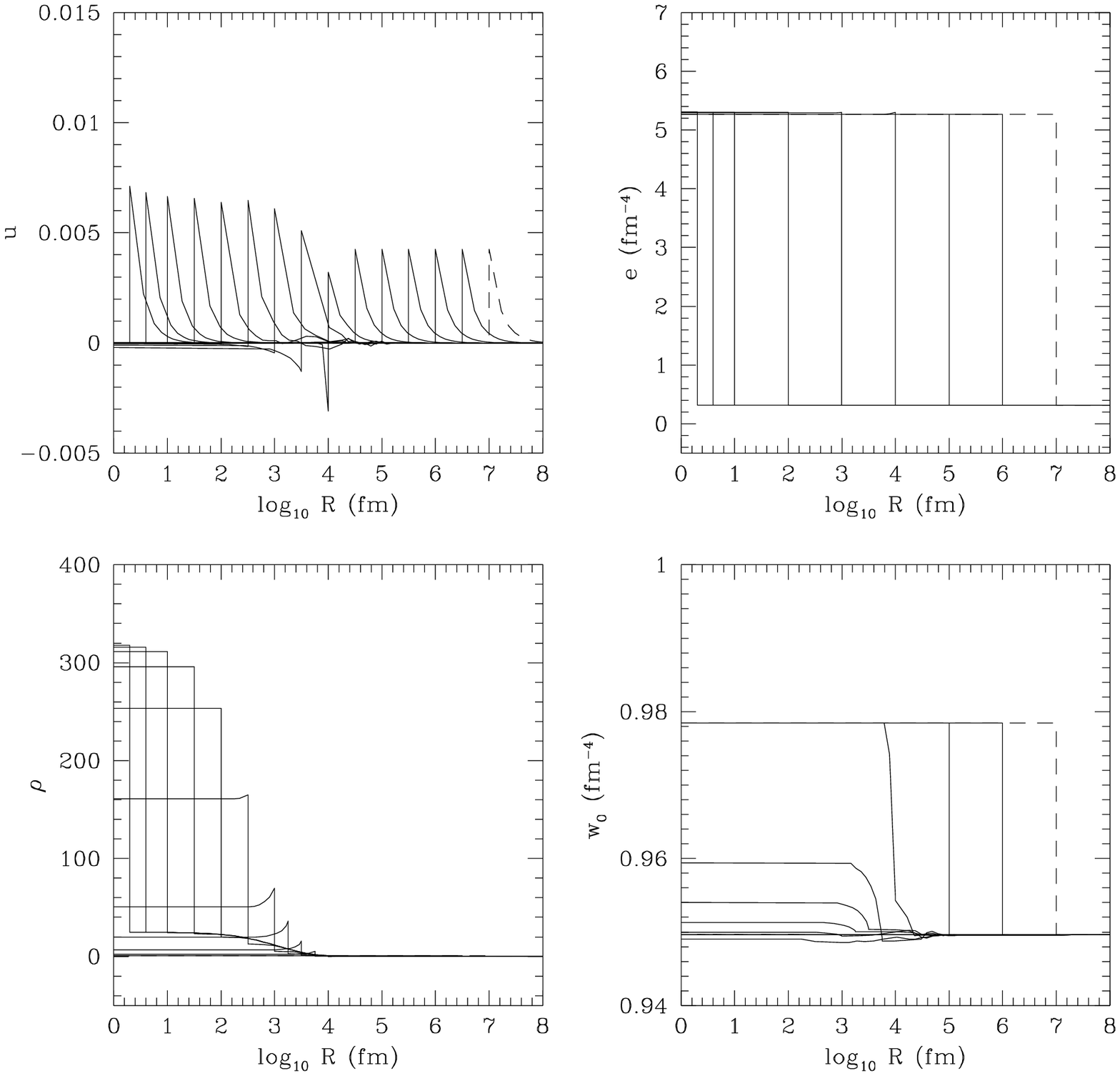,height=16.0truecm,width=17.0truecm}}}
\medskip
\vskip 0.5truecm
\centerline{\vbox{\hsize 12.0truecm\baselineskip=12pt\noindent\small
Figure 1. Time evolution of the most significant hydrodynamical variables. 
Each curve refers to a given time in the solution, with the quark phase
being to the left of the vertical discontinuity and the initial conditions
being indicated with the dashed curves. Starting from the upper left-hand
window and proceeding clockwise, the frames show: the radial component
of the fluid four-velocity in the Eulerian frame $u$, the energy
densities of the standard fluids $e$ and of the radiation fluid $w_0$
and the compression factor $\rho$.  The decoupling between the
radiation fluid and the standard fluids is allowed to start at
$R_s=10^4$ fm} } }
\bigskip

\vsize=24.0truecm 

\bigskip \vbox{\vskip -2.5truecm \hskip -1.75truecm 
\centerline{{\psfig{figure=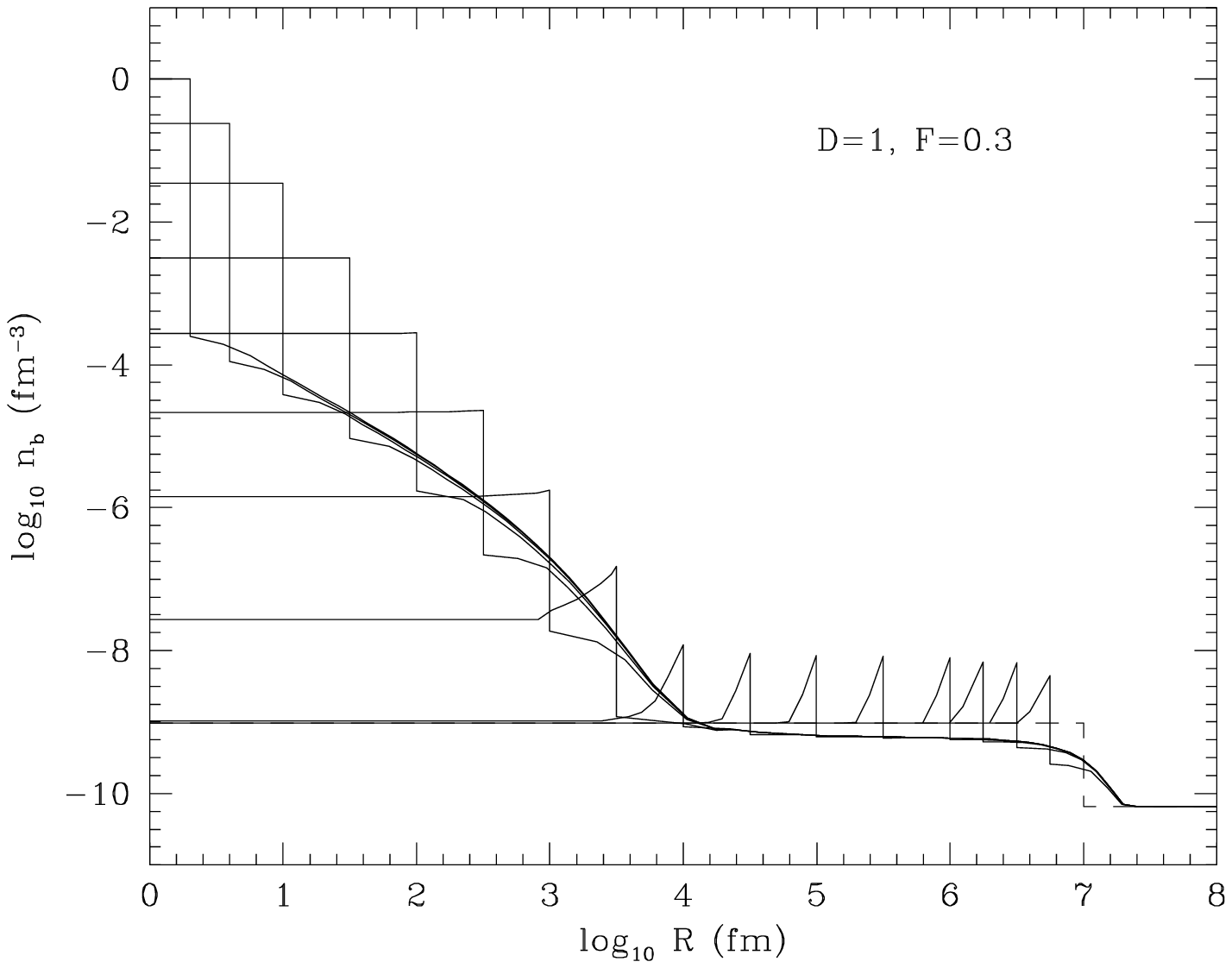,height=8.0truecm,width=10.0truecm}}} 
\medskip \vskip 0.5truecm 
\centerline{\vbox{\hsize 12.0truecm\baselineskip=12pt\noindent\small 
 Figure 2. Time evolution of the logarithm of the baryon number density. 
The continuous curves refer to successive stages in the evolution, with
the initial conditions being indicated by the dashed curve. 
 } }} \medskip

\medskip
\vbox{\vskip -0.0truecm \hskip -1.15truecm \centerline{
{\psfig{figure=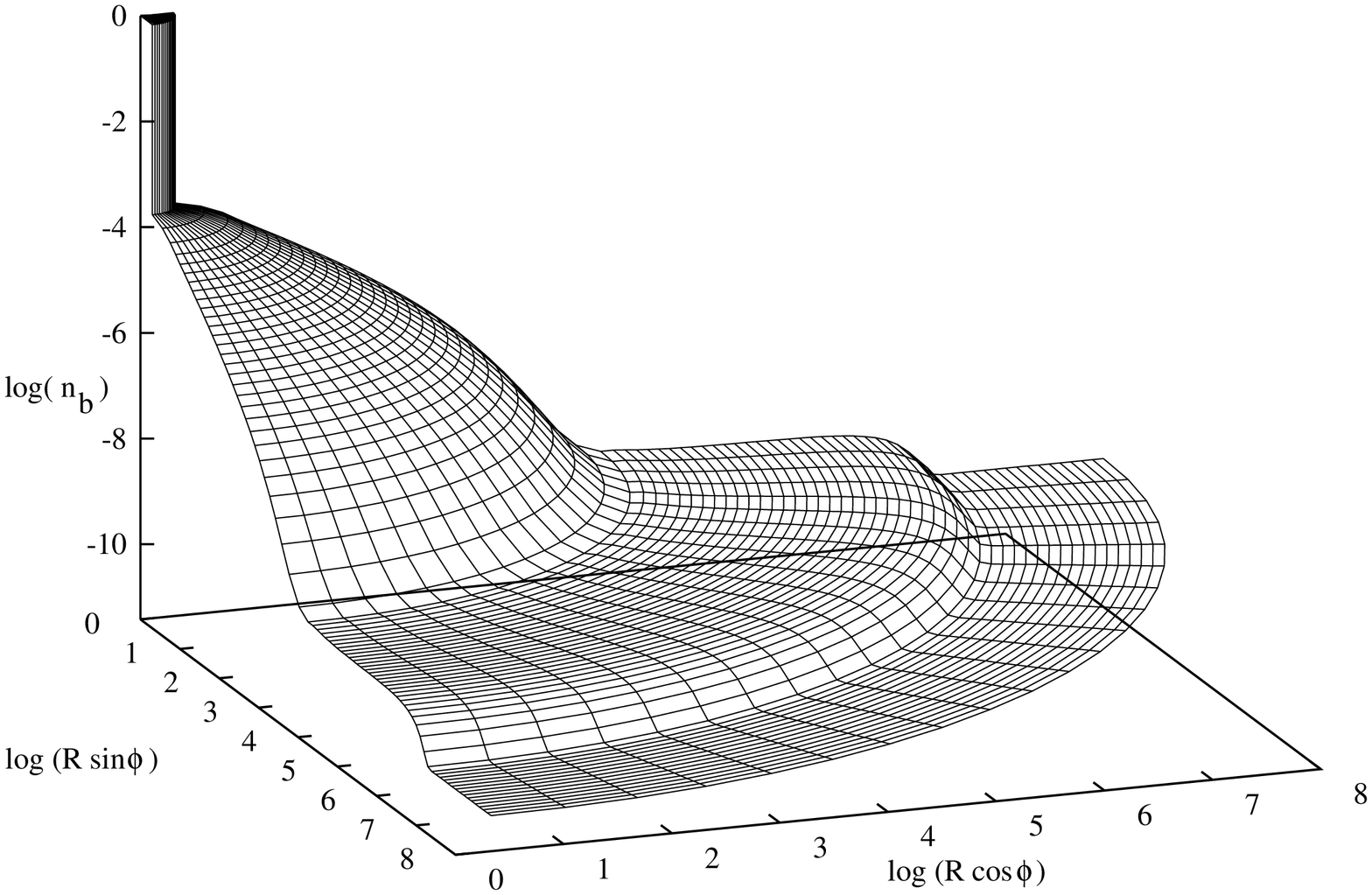,height=9.5truecm,width=14.5truecm,angle=270}}}
\medskip
\vskip 0.25truecm
\centerline{\vbox{\hsize 12.0truecm\baselineskip=12pt\noindent\small
Figure 3. Three-dimensional plot of the profile of the log of the
baryon number density at the end of the computation, when the quark
drop had a radius $R_s=2$ fm.}} }
\vfill\eject

\vsize=20.0truecm 

\noindent
 This decoupling would take place at a scale $\lambda_{\nu} \simeq
10^{13}$ fm but it is not currently clear whether spherically isolated
quark drops would have been present at that scale.
	
	Figure 1 shows the time evolution of the most important
hydrodynamical variables: the radial component of the fluid
four-velocity $u$, the energy density of the standard fluids $e$, the
compression factor $\rho$ and the energy density of the radiation fluid
$w_0$. Although similar curves have already been presented and
discussed in \cite{rm96}, we consider it useful to show them here as
they provide the hydrodynamical background for our discussion of baryon
number segregation.

	The main features of the hydrodynamical solution can be summarized
as follows: {\it i)} the presence of a self similar solution until the
radiation decoupling starts at $R_s=R_d$; {\it ii)} the recovering of
an ``almost'' self similar solution for the energy density $e$ and the
fluid velocity $u$ after decoupling; {\it iii)} the significant
increase of the compression factor in both phases caused by the
extraction of entropy by the radiation fluid; {\it iv)} the smoothing
out of the step in the profile of the radiation energy density when
decoupling occurs.
	
	In Figure 2 we show the time evolution of the baryon number
density which, for convenience, we can divide into three main stages.
In the first of these, baryon number density deviates from the imposed
initial conditions and settles into a solution which is consistent with
the motion of the interface. As mentioned above, the initial conditions
adopted assume chemical equilibrium to hold across the phase interface.
However, when the interface is in motion (as it is for our initial
conditions), equilibrium cannot hold and the baryon number density in
the hadron phase has to adjust itself in order to reach a
self consistent solution. By the time the drop radius has decreased
from $10^7$ fm to approximately $2.5 \times 10^6$ fm, the new solution
has reached its steady state configuration.

	The next stage of the evolution is characterized by a ``snowplow''
behaviour of the phase interface which accumulates baryon number as it
moves into the quark medium. This piling-up of baryon number on the quark
side is counteracted by the effects of diffusion which tends to smear out
the build-up, producing a rapidly decaying profile (a similar behaviour
was also suggested by Kurki-Suonio \cite{ks88}). During this stage, which
corresponds to a self similar solution for the hydrodynamical variables
presented in Fig. 1, the counteracting mechanisms of baryon number flux
suppression and of baryon diffusion, suitably regulate the rate of baryon
number passage across the phase interface until a stationary solution is
reached for the baryon number density. Note that the length scale $r_d$ is
too small to be resolved on the grid at this stage and this explains why,
in Fig. 2, the decaying profile appears to have a constant logarithmic
width (connected with the grid structure). 

	The self similar solution for the hydrodynamical variables can
be maintained only as long as the quark drop evaporation is
effectively scale free and no other process is intervening.
Equivalently, the stationary solution for baryon number can be
maintained only as long as the underlying hydrodynamical behaviour
remains self similar and the tail of the diffusing profile does not
interact with the centre of the drop (the time when this happens is
dependent on the magnitude of $D$). In the present situation there are
two possible length scales that when reached by the drop radius would
produce a deviation away from the hydrodynamic similarity
solution. These are the mean free path of the radiation fluid
particles (at which scale the radiative transfer is most effective)
and the length scale set by the surface tension (which is much
smaller).

	For our fiducial parameter values, the radiative transfer occurs
before the drop can know about its centre via diffusion and, as a
consequence, the following stage of the evolution in Fig. 2 starts at
$R_s=10^4$ fm. When we compare the effects of the entropy extraction via
radiative transfer on the compression factor (lower-right window in Fig.
1) and on the baryon number density, we can find analogies and
differences. The first common feature is the increase of $\rho$ (and of
$n_b$) {\it also} in the low temperature phase for $R_s < 10^4$ fm.
However, while the radius at which the compression increase in the hadron
phase begins does not change in time, this is not the case for the radius
at which the increase in baryon number density begins. The latter slowly
shifts towards larger values under the action of baryon diffusion. Figure
3 shows a three-dimensional representation of the profile of baryon number
density at the end of the computation when the quark drop has radius
$R_s=2$ fm.  \footnote{ It is important to stress that, we here stop our
computations at $R_s=2$ fm because already at this stage the whole
hydrodynamical approach described in Sections I--III is ceasing to be
valid. It is no longer sensible to use a {\it fluid} description for a
quark drop containing only very few quarks.}
 
	 Another difference between the evolution of the compression
factor and that of the baryon number density is the larger increase in
the latter (about 9 orders of magnitude) produced after radiation
decoupling. This is due to the fact that baryon number density can be
increased not only by the entropy extraction via radiative transfer
but also by flux suppression at the interface and the latter is a much
more effective mechanism. Note that the profile of baryon number
density shown in Fig. 3 corresponds to material which has non-uniform
specific entropy (as a result of radiative transfer \cite{rm96}), but
which is very nearly in thermal and mechanical equilibrium in each
phase \cite{jfmk94,ah85}.

	When looking at the profiles in Figures 2 and 3 it is not possible
to distinguish the different contributions coming from the radiative
transfer and from the flux suppression. However, using the mean free path
of the radiation particles $\lambda$ as a free parameter, it is possible
to regulate the action of the radiative transfer.  Taking $\lambda=0$
corresponds to a situation in which the decoupling never takes place at
all. Figure 4 shows the final profile of the baryon number density when
the quark drop has radius $R_s=2$ fm as calculated for progressively
smaller values of the mean free path, which is equal to $\lambda=10^4$ fm
for the heavy continuous curve, $\lambda=10^2$ fm for the dashed curve and
to zero for the dotted one. 

	The effects of radiative transfer can be clearly seen by comparing
the dotted curve with the heavy continuous curve and amount to a relative
increase in the baryon number density of around one order of magnitude in
the inner part of the profile. Note that the use of a logarithmic scale
can be misleading and seem to suggest that total baryon number in the
different plots is not the same. However, when comparing profiles for
different values of $\lambda$, one should pay particular attention to
differences in the outer regions which are not very noticeable but which
make a considerable contribution. In order to emphasize these differences,
a magnification of the final profiles is shown in the small diagram at the
top right of Figure 4.
	
\bigskip
\vbox{\hskip -1.75 truecm \centerline{
{\psfig{figure=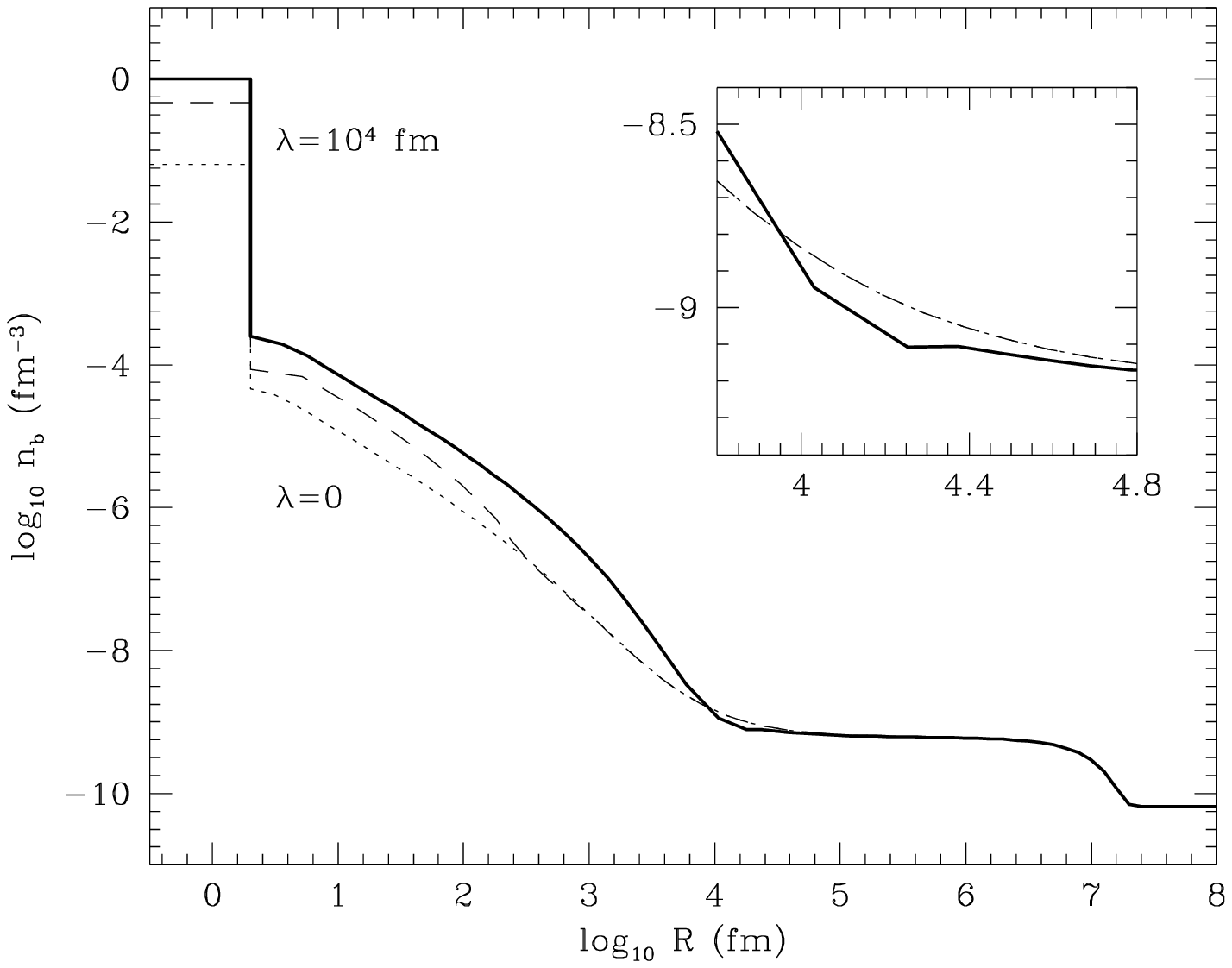,height=8.0truecm,width=10.0truecm}}}
\medskip
\vskip 0.5truecm
\centerline{\vbox{\hsize 12.0truecm\baselineskip=12pt\noindent\small
Figure 4. Final profile of the baryon number density when the quark
drop has radius $R_s=2$ fm as calculated for different values of the mean 
free path $\lambda$. The curves correspond to $\lambda=10^4$ fm (heavy
continuous curve), $10^2$ fm (dashed curve) and $0$ (dotted curve).
The small diagram at the top right shows a magnification of the
differences between the curves.} } }
\bigskip 
\medskip

	Figure 5 shows results from calculations performed with different
values of the filter factor, showing the final profiles of baryon
number density when the drop has radius 2 fm. Different curves refer to
values of the filter factor $F$ ranging from $0.2$ to $1$, with the
heavy continuous curve corresponding to the preferred value $F=0.3$.

	The behaviour of the solution under variation of the filter
factor $F$ is straightforward to understand: a {\it more}
``transparent'' phase interface (i.e. with a higher value of $F$)
leads to less accumulation of baryon number inside the drop and
produces a smaller final increase in the baryon number
density. However, because of the nonlinearity of the process, a {\it
less} transparent interface actually gives a larger net baryon number
flux as a consequence of the increased baryon number density in the
quark phase [see equation (\ref{phib})]. This explains why the baryon
number density is larger also outside the quark drop. Note that also
in this figure, the logarithmic scaling could be confusing and suggest
that the total baryon number is not the same for the different curves.
However, once again, as shown in the small magnified diagram, the
large differences which appear for $R_s < 10^4$ fm are compensated by
changes further out which are less easy to see.

\bigskip
\vbox{\hskip -1.75 truecm \centerline{
{\psfig{figure=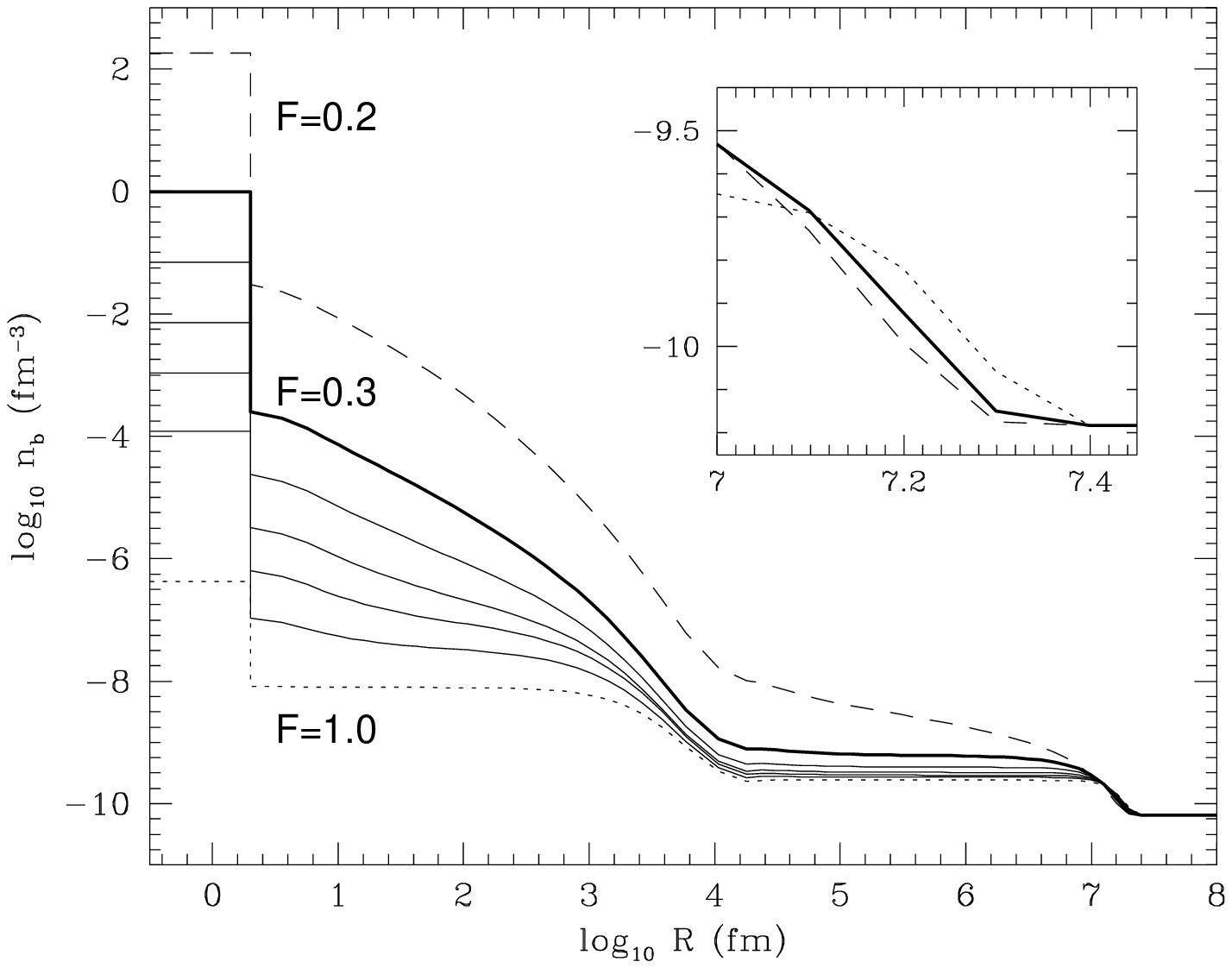,height=8.0truecm,width=10.0truecm}}}
\medskip
\vskip 0.5truecm
\centerline{\vbox{\hsize 12.0truecm\baselineskip=12pt\noindent\small
 Figure 5. Final profile of the baryon number density when the quark
drop has radius $R_s=2$ fm as calculated for different values of the
filter factor at the interface $F$. The curves correspond to $F=0.2$
(dashed curve), $F=0.3$ (heavy continuous curve), $F=0.4,\ 0.5,\ 0.6,\
0.7$ (standard continuous curves) and $F=1$ (dotted curve). The
small diagram at the top right shows a magnification of the
differences between the most important curves.} } }
\bigskip \medskip

	A number of comments should be made about the final baryon
number density reached inside the quark phase for different values of
the filter factor. As shown by the curve corresponding to $F=0.2$, the
baryon number density in the quark phase could well reach values above
that for nuclear matter and this could suggest that a transition to
strange quark nuggets might possibly occur at some stage during the
quark drop evaporation. This is a very speculative idea and we limit
ourselves to pointing out that, as far as baryon number segregation is
concerned, conditions for the creation of quark nuggets are not
difficult to reach if a strong suppression of baryon number flux takes
place at the interface. Another important point to notice is that with
the present choice of parameter values, only a very small fraction of
the initial total baryon number will remain in the high density
regions after drops have evaporated away. A final comment should be
made about the properties of the final baryon number density profile
obtained for $F=1$, which corresponds to a situation in which baryon
number carriers are freely streaming across the interface and is
the result of radiation decoupling only. Although this case may well
not refer to a realistic scenario, it is instructive as it shows the
role of the baryon flux suppression in the production of baryon number
density peaks. It is interesting to compare this with the $\lambda=0$
curve of Fig. 4 which corresponds to radiative transfer not being
operative.

	We conclude our analysis of the numerical results by presenting
the final baryon number density profiles obtained after varying the
diffusion coefficient while maintaining the other parameters with their
standard values. Figure 6 shows curves calculated for values of $D$
ranging between zero (dotted curve) and 10 and with the heavy continuous
curve corresponding to the preferred value $D=1$. 

	The curves demonstrate the action of diffusion which prevents the
accumulation of baryon number in a narrow shell ahead of the phase
interface (as seen for $D=0$) and spreads out the accumulation of baryon
number produced in the hadron phase near to the drop surface. Larger
values of the diffusion coefficient give rise to larger final values of
the baryon number density in the quark phase, but the results show that
the high number density regions are unlikely to contain a large fraction
of the total baryon number if only microscopic random diffusion is taken
into account (which leads to relatively small values for $D$).  

\bigskip
\vbox{\hskip -1.75 truecm \centerline{
{\psfig{figure=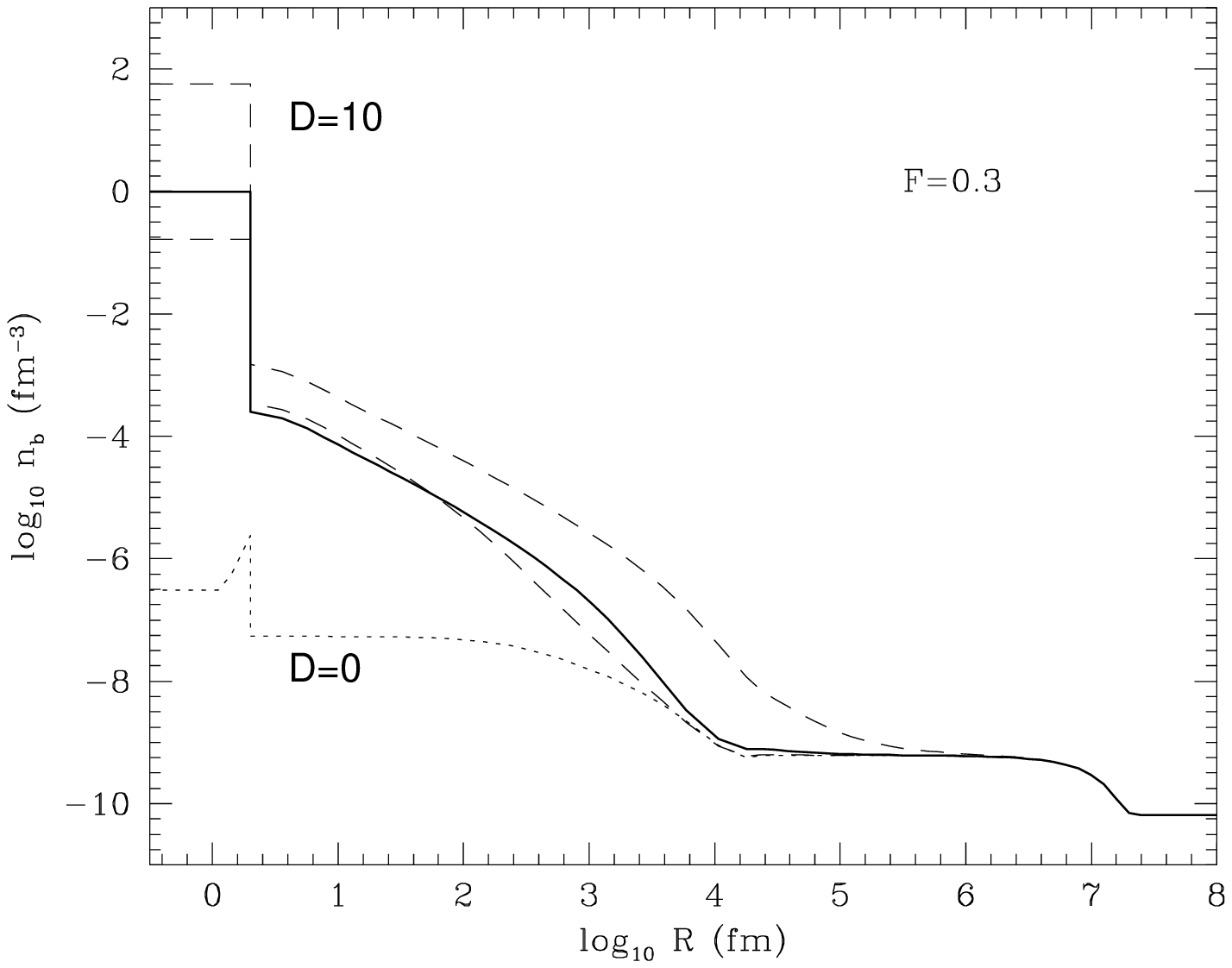,height=8.0truecm,width=10.0truecm}}}
\medskip
\vskip 0.5truecm
\centerline{\vbox{\hsize 12.0truecm\baselineskip=12pt\noindent\small
Figure 6. Final profile of the baryon number density when the quark
drop has radius $R_s=2$ fm as calculated for different values of the
diffusion coefficient $D$. The curves correspond to $D=10,\ 1$ (the
heavy continuous line), $0.1$ and $0$ (the dotted curve).} } }
\bigskip
\medskip

	Stronger diffusion is more efficient in preventing accumulation of
baryon number adjacent to the quark side of the interface and this, in
turn, tends to reduce the outward flux of baryon number into the hadron
phase. The effective role played by diffusion is then that of trapping
baryon number within the high density region and so it is no surprise that
higher final values of $n_b$ should result for larger values of $D$.  It
is worth underlining that the values adopted for our simulations refer to
a microscopic modeling of diffusive processes in which turbulent motion is
neglected. It is possible that non radial fluid motions could produce a
larger effective diffusion coefficient, but not enough is yet known about
this. 

{\who=1\bigskip\medskip\bigskip\goodbreak
{\Large\bf\noindent\hbox{VI.}\hskip 0.5truecm Conclusions }\nobreak
\bigskip\medskip\nobreak\who=0}

	We have presented here results from numerical computations of the
final stages of a first order cosmological quark--hadron phase
transition, aiming to estimate the baryon number density profile
produced. For doing this, we have used mathematical and numerical
techniques previously developed for following the evaporation of
spherical quark drops together with a separate treatment of a distinct
fluid of baryon number carriers. The effects of baryon number flux
suppression at the phase interface and of baryon number diffusion have
been included within the overall relativistic hydrodynamical treatment.
The results obtained have shown that baryon number segregation is an
inevitable result of a first order transition and illustrate the
different roles played in baryon number segregation by flux suppression
and by radiative transfer. Final baryon number density profiles
have been computed which exhibit relative increases of several orders of
magnitude in both phases, with an overdense spherical region of radius
\hbox{$R\approx \lambda \approx 10^4$} fm being left behind in thermal
equilibrium in the hadron phase. This region, which probably
represents the most important relic of the transition, will be subject
to further subsequent diffusion.

 	As pointed out by Ignatius et al. \cite{ikksl94b}, a number of
conditions would need to be met if baryon number segregation is to
have any significant effect on nucleosynthesis. One of these
conditions (that the final baryon number density contrast should be
larger than $\sim 10$) is certainly satisfied by the results from our
present study. However, our results also indicate that the condition
for the high density regions to contain most of the total baryon
number is extremely difficult to satisfy unless turbulent diffusion
might be operative. A third necessary condition requires that the mean
separation scale $l_n$ between two baryon number density peaks of the
type computed here should be rather large. It would be necessary to
have $l_n > 1\ {\rm m} \approx 10^{-4}\; t_H$, where $t_H$ is the
horizon scale at the time of the transition. As mentioned in the
Introduction, this is a fundamental quantity for the whole description
of the phase transition, without which any conclusion about the role
played by baryon number inhomogeneities on the subsequent
nucleosynthesis remains extremely unclear. Considerable efforts have
been devoted to determining it (see Christiansen and Madsen
\cite{cm96} for a recent discussion) but the value of this scale is
still uncertain and further developments need to be awaited.

	The hydrodynamical scenario for baryon number segregation
discussed here, could represent a promising starting point for studying
the possible production of seed magnetic fields during the quark--hadron
phase transition. Cheng and Olinto \cite{co94} pointed out that because of
the different fractional electric charge carried by up and down quarks,
charge separation could be produced together with the concentration of
baryon number in the high temperature phase. In their paper, they focussed
attention on the slow growth stage of the phase transition during which
chemical equilibrium holds and the baryon number contrast can be estimated
analytically. However, this same mechanism might be at work also during
the final stages of the transition, for which the present calculations
provide more precise quantitative estimates of the dynamical motions of
charges and of their density. 


	Because of the much more extreme baryon number concentrations
arising in the present context, the magnetic field could be several orders
of magnitude larger than that discussed in \cite{co94} and might possibly
reach the equipartition value. This requires further investigation and
will be a focus of future work. 

\bigskip
\bigskip
{\bigskip\bigskip\medskip\noindent
\Large\bf{\hskip 0.5 truecm Acknowledgments}\bigskip}

        It is a pleasure to thank John C. Miller for his guiding comments,
for suggesting the convenient finite difference expression (\ref{fntdif})
and for carefully reading this manuscript.

\vfill\eject

{\who=1\bigskip\medskip\bigskip\goodbreak
{\Large\bf\noindent\hbox{   }\hskip 0.5truecm Appendix }\nobreak
\bigskip\medskip\nobreak\who=0}

	In this Appendix, we briefly sketch some intermediate steps leading
to equation (\ref{cont_nb1}), the relativistic diffusion equation for
baryon number. Using the standard property of four-divergences
mentioned in the text, equation (\ref{cont_nb}) can be written as

\begin{equation} 
\label{a1}
b R^2\left[ (n_b)_{,\;t} +
n_b\left( {b_{,\;t}\over b} + {2 R_{,\;t}\over R} \right) \right]=
-a b \left[ 
R^2q^{\mu} \left({a_{,\;\mu}\over a}+{b_{,\;\mu}\over b} \right)
+ (R^2 q^{\mu}) \right] \ .
\end{equation}

\noindent
The $T^0_{\ 1}$ component of the Einstein field equation provides the
useful relation

\begin{equation} 
\label{a2}
{b_{,\;t}\over b} = {R_{,\;\mu t}\over R_{,\;\mu}}  -
{a_{,\;\mu}\over {a R_{,\;\mu}}} -
4 \pi G R a b w_1 \approx
{R_{,\;\mu t}\over R_{,\;\mu}}  -
{a_{,\;\mu}\over {aR_{,\;\mu}}} =
{a \over {b \Gamma}} u_{,\;\mu}\ ,
\end{equation}

\noindent
where $G \ll 1$ is the gravitational constant and we drop the term
containing it for the present purposes. Making use of the identity

\begin{equation} 
\label{a3}
{b_{,\;\mu}\over b} = 
{R_{,\;\mu \mu}\over {b \Gamma}}  -
{\Gamma_{,\;\mu}\over {b \Gamma}}\ ,
\end{equation}

\noindent
 and of the definition (\ref{qmu}), it is then possible to rewrite
equation (\ref{a1}) in the final form (\ref{cont_nb1}). 

\vfill\eject

\end{document}